\documentclass[journal,twoside,10pt]{IEEEtran}

\usepackage{amsmath,amssymb,amsthm}
\usepackage{algorithm}
\usepackage{algpseudocode}          % algorithmicx / algpseudocode
\usepackage{graphicx}
\usepackage{cite}
\usepackage{url}
\usepackage{booktabs}
\usepackage{multirow}
\usepackage{xcolor}
\usepackage{bm}
\usepackage{tikz} 
\usetikzlibrary{shapes.geometric, arrows.meta, positioning, calc}
\usepackage{comment}
\usepackage[font=footnotesize,labelfont=bf,labelsep=period]{caption}

\usepackage{subcaption}
\usepackage{multirow}
\usepackage{threeparttable}
\usepackage{hyperref}
\usepackage{siunitx}
\usepackage{orcidlink}

\usepackage{soul}
\usepackage{xcolor}

\usepackage{tikz}

\definecolor{c0}{RGB}{0,114,189}
\definecolor{c1}{RGB}{217,83,25}
\definecolor{c2}{RGB}{237,177,32}
\definecolor{c3}{RGB}{126,47,142}
\definecolor{c4}{RGB}{119,172,48}
\definecolor{c5}{RGB}{77,190,238}

\algrenewcommand\algorithmicrequire{\textbf{Input:}}
\algrenewcommand\algorithmicensure{\textbf{Output:}}
\algnewcommand{\LineComment}[1]{\State \(\triangleright\) \textit{#1}}

\begin{document}

\title{Deep Learning-Assisted Optimization of Large Domino-Tiled Phased Arrays for Dual-Band LEO Satellite Communications}

%\author{Mohammad~Mashayekhi,~\IEEEmembership{Member,~IEEE,}
   %     Author~Two,~\IEEEmembership{Senior Member,~IEEE,}
   %     and~Author~Three,~\IEEEmembership{Member,~IEEE}%

\author{Mohammad~Mashayekhi\,\orcidlink{0009-0004-4865-1264},
	Gholamreza~Askari\,\orcidlink{0000-0003-3706-2734}

\thanks{Manuscript received Month DD, YYYY; revised Month DD, YYYY.}

\thanks{The authors are with the Information and Communication Technology Institute (ICTI), Isfahan University of Technology, Isfahan 84156-83111, Iran (e-mail:askarigh@iut.ac.ir; .}
}

\markboth{IEEE TRANSACTIONS ON ANTENNAS AND PROPAGATION,~VOL.~XX, NO.~X, MONTH~YEAR}%
{Mashayekhi \MakeLowercase{\textit{et al.}}: Deep Learning-Assisted Optimization of Large Domino-Tiled Phased Arrays }
 
\maketitle

\begin{center}
	\footnotesize
	\textit{This manuscript has been submitted to the IEEE
		Transactions on Antennas and Propagation for possible
		publication and is currently under review.}
\end{center}

\begin{abstract}
	Irregular domino tiling reduces the number of transmit/receive modules (TRMs) of large phased arrays, but its design requires repeated pattern evaluations whose cost grows rapidly with aperture size. This paper presents a hybrid deep learning and optimization framework for large domino-tiled arrays. A local-flip Markov chain generator, about $5.2\times$ faster than a randomized greedy baseline, produces balanced tilings with nearly equal numbers of horizontal and vertical dominoes, and a Tabu Search engine reaches a sidelobe level (SLL) about $0.8$~dB lower than a genetic algorithm, simulated annealing, and iterated local search at comparable directivity. These tools build the dataset on which a convolutional neural network (CNN) surrogate is trained to predict, from the tiling structure alone, whether a candidate yields a favorable multi-angle SLL. Gating the Tabu Search with the surrogate removes $60.6$--$70.6\%$ of the full evaluations and speeds up the search by $2.56$--$3.80$ times for apertures from $20\times20$ to $100\times100$ elements, without degrading the solution quality. A convex amplitude--phase optimization of the fixed tilings then gives equal or higher directivity than two state-of-the-art methods at a comparable SLL, with more than an order of magnitude shorter computation time. Finally, the framework is applied to a $36\times36$ dual-band circularly polarized array for low Earth orbit (LEO) terminals at $11.2$ and $14.0$~GHz, whose bands are a factor of $1.25$ apart, using $648$ dominoes and a common element spacing selected from a sweep. The array scans to $50^\circ$ with no grating lobe and, compared with a reference irregular array, gives higher directivity over most of the scan range, a lower SLL at large scan angles, and a spacing that is more practical for the lower band.
\end{abstract}


\section{Introduction}
\label{sec:intro}

Phased array antennas~(PAAs) are widely used in radar, fifth-generation~(5G) and beyond
communications, satellite links, autonomous vehicles, and terahertz imaging
\cite{alhamed2023, chen2021, ku2014, chen2022satellite}. In conventional phased
arrays, each element is driven by a dedicated transmit/receive~(T/R) module, which alone accounts
for nearly half the system cost~\cite{herd2016}. As apertures grow to thousands of elements, this
per-element architecture becomes prohibitively expensive to build and maintain~\cite{Zhao2021}.

\begin{figure}[t]
	\centering
	\includegraphics[width=0.8\linewidth]{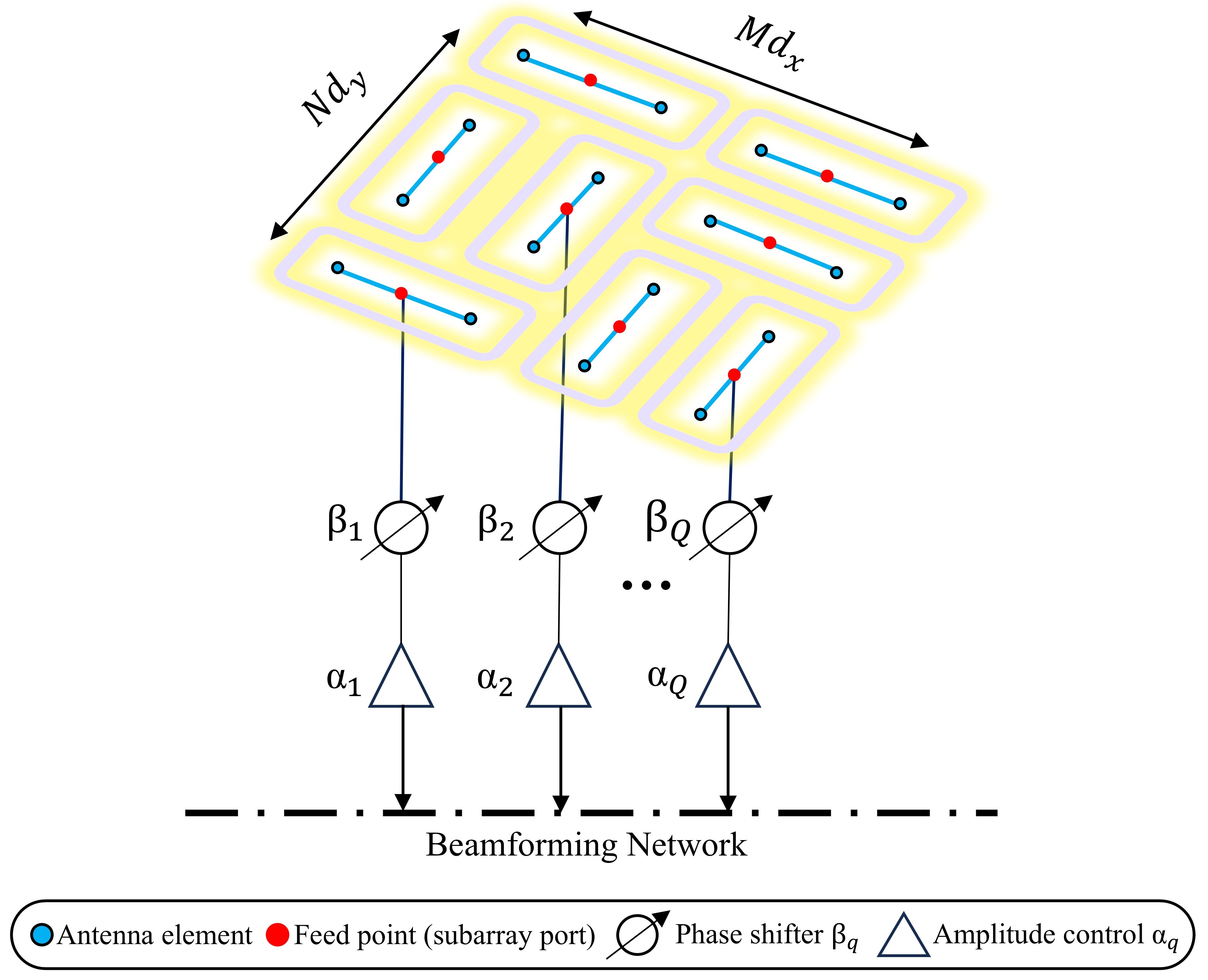}
	\caption{Architecture of a domino-tiled phased array antenna.}
	\label{fig:tiled_array}
\end{figure}

Thinned~\cite{Haupt1994, Salucci2017}, sparse~\cite{Bucci2013, Bencivenni2016}, and tiled arrays
have been investigated as cost-effective alternatives. Thinned and sparse arrays suffer from low
aperture efficiency, which limits their gain~\cite{rocca2020,rocca2016review}. In tiled arrays,
the aperture is instead partitioned into non-overlapping polyomino-shaped subarrays
\cite{mailloux2009, karademir2016, rocca2015ga, anselmi2017, ma2021directivity,
	yang2021convex, li2021hierarchical, liu2023sctie, chen2023sidelobe,
	jiang2023algox, Ma2024, chen2025novel, jin2023hierarchical, yang2024tiling,
	ertugrul2025triangular, rocca2016review}, each driven by a single T/R module
(Fig.~\ref{fig:tiled_array}~\cite{mou2022repetitiveness}). This preserves aperture efficiency,
makes hardware cost scale with the number of subarrays, and simplifies manufacturing and
calibration~\cite{li2021hierarchical, yang2024tiling}. Irregular tilings, which use non-uniform
subarray shapes, are more effective at suppressing grating lobes than regular ones
\cite{mailloux2009}.

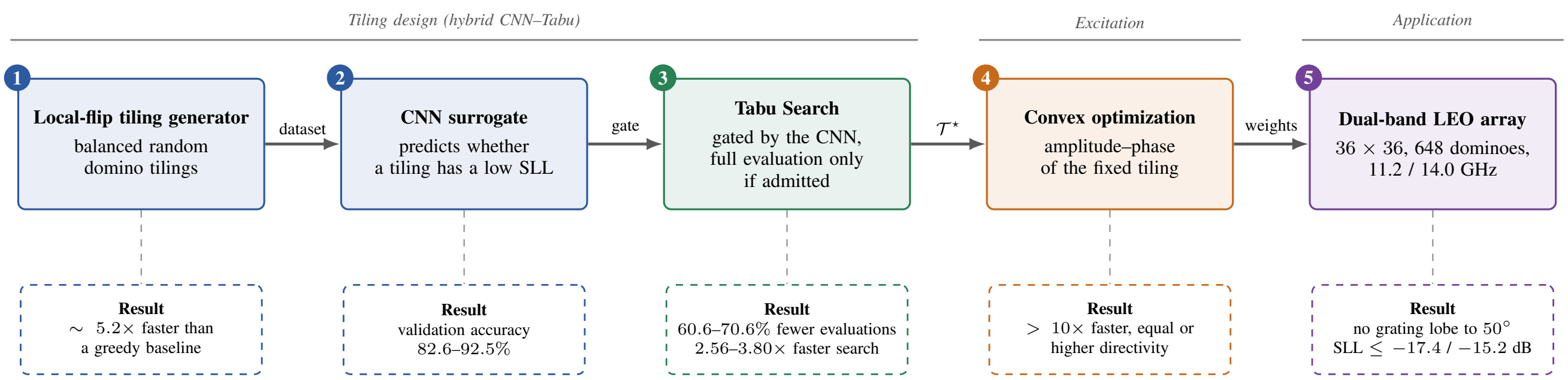
\begin{figure*}[!t]
	\centering
	\definecolor{cA}{RGB}{44,90,160}
	\definecolor{cB}{RGB}{40,130,85}
	\definecolor{cC}{RGB}{200,105,25}
	\definecolor{cD}{RGB}{115,65,155}
	\resizebox{0.95\textwidth}{!}{%
		\begin{tikzpicture}[
			font=\footnotesize,
			box/.style={rectangle, draw=#1, fill=#1!10, rounded corners=3pt,
				text width=3.3cm, align=center, minimum height=1.9cm, inner sep=4pt, line width=0.8pt},
			num/.style={circle, fill=#1, text=white, inner sep=1.5pt, font=\footnotesize\bfseries},
			arr/.style={-{Latex[length=2.4mm]}, line width=1.2pt, draw=black!65},
			grp/.style={line width=0.8pt, draw=black!45},
			glab/.style={font=\scriptsize\itshape, text=black!65, above=1pt},
			res/.style={rectangle, rounded corners=3pt, draw=#1, dashed, fill=white, font=\scriptsize,
				text width=3.3cm, align=center, minimum height=1.3cm, inner sep=3pt, line width=0.7pt}
			]
			
			\node[box=cA] (n1) at (0,0)    {\textbf{Local-flip tiling generator}\\[3pt] balanced random\\ domino tilings};
			\node[box=cA] (n2) at (4.7,0)  {\textbf{CNN surrogate}\\[3pt] predicts whether\\ a tiling has a low SLL};
			\node[box=cB] (n3) at (9.4,0)  {\textbf{Tabu Search}\\[3pt] gated by the CNN,\\ full evaluation only\\ if admitted};
			\node[box=cC] (n4) at (14.1,0) {\textbf{Convex optimization}\\[3pt] amplitude--phase\\ of the fixed tiling};
			\node[box=cD] (n5) at (18.8,0) {\textbf{Dual-band LEO array}\\[3pt] $36\times36$, 648 dominoes,\\ 11.2 / 14.0~GHz};
			
			\node[num=cA] at (n1.north west) {1};
			\node[num=cA] at (n2.north west) {2};
			\node[num=cB] at (n3.north west) {3};
			\node[num=cC] at (n4.north west) {4};
			\node[num=cD] at (n5.north west) {5};
			
			\draw[arr] (n1) -- node[above, font=\scriptsize]{dataset} (n2);
			\draw[arr] (n2) -- node[above, font=\scriptsize]{gate} (n3);
			\draw[arr] (n3) -- node[above, font=\scriptsize]{$\mathcal{T}^\star$} (n4);
			\draw[arr] (n4) -- node[above, font=\scriptsize]{weights} (n5);
			
			% group brackets
			\draw[grp] ($(n1.north west)+(-0.1,0.55)$) -- ($(n3.north east)+(0.1,0.55)$)
			node[glab, midway]{Tiling design (hybrid CNN--Tabu)};
			\draw[grp] ($(n4.north west)+(-0.1,0.55)$) -- ($(n4.north east)+(0.1,0.55)$)
			node[glab, midway]{Excitation};
			\draw[grp] ($(n5.north west)+(-0.1,0.55)$) -- ($(n5.north east)+(0.1,0.55)$)
			node[glab, midway]{Application};
			
			% results
			\node[res=cA] (r1) at (0,-2.7)    {\textbf{Result}\\ $\sim5.2\times$ faster than\\ a greedy baseline};
			\node[res=cA] (r2) at (4.7,-2.7)  {\textbf{Result}\\ validation accuracy\\ $82.6$--$92.5\%$};
			\node[res=cB] (r3) at (9.4,-2.7)  {\textbf{Result}\\ $60.6$--$70.6\%$ fewer evaluations\\ $2.56$--$3.80\times$ faster search};
			\node[res=cC] (r4) at (14.1,-2.7) {\textbf{Result}\\ $>10\times$ faster, equal or\\ higher directivity};
			\node[res=cD] (r5) at (18.8,-2.7) {\textbf{Result}\\ no grating lobe to $50^\circ$\\ SLL $\leq-17.4$ / $-15.2$~dB};
			\foreach \i in {1,...,5}{\draw[dashed, black!40, line width=0.6pt] (n\i.south) -- (r\i.north);}
			
	\end{tikzpicture}}
	\caption{Overview of the proposed framework: (1)~a local-flip generator produces balanced domino tilings; (2)~a CNN trained on them predicts whether a tiling has a favorable SLL; (3)~it gates the full evaluations of a Tabu Search that returns the tiling $\mathcal{T}^\star$; (4)~a convex amplitude--phase optimization gives the final excitations; (5)~the framework is applied to a dual-band circularly polarized LEO array.}
	\label{fig:overview}
\end{figure*}

Irregular tiling requires jointly determining the aperture partition and the subarray
excitations under radiation constraints, a combinatorially complex problem. Genetic algorithms
based on analytic tileability theorems \cite{anselmi2017}, a maximum-entropy formulation
\cite{ma2021directivity}, and a weighted $\ell_1$-norm convex relaxation \cite{yang2021convex}
have been proposed. For larger apertures, a hybrid of deterministic tiling and evolutionary
search \cite{yang2024tiling} and a two-level hierarchical optimization \cite{jin2023hierarchical}
reduce the computational cost. Alternatively, structural constraints such as replication and
flipping of a single subarray \cite{mou2022repetitiveness}, rotational symmetry
\cite{chen2023sidelobe}, or hierarchical twice-partitioning \cite{jiang2023algox} simplify the
design, but at the price of limited pattern quality compared with fully irregular tilings.
Despite their differences, all unconstrained approaches rely on repeated radiation-pattern
evaluations of each candidate, which become the dominant bottleneck as the aperture and the
number of iterations grow.
Recent works point toward larger apertures~\cite{yang2024tiling,Anselmi2026} and data-driven
synthesis~\cite{Ma2025}. The tiling problem, with its exponentially large solution space, is a
natural fit for deep learning, which has already shown strong potential in electromagnetic
design~\cite{Mashayekhi2023,Mashayekhi2025DL,Mashayekhi2025SIW,Ghorbani2021}.
This paper proposes a deep-learning-assisted framework (Fig.~\ref{fig:overview}). A fast
local-flip generator and a Tabu Search engine are used to build a labeled dataset, on which a
convolutional neural network~(CNN) surrogate is trained to predict, directly from the tiling
structure, whether the multi-angle sidelobe level~(SLL) of a candidate is favorable. The surrogate
then gates the array-factor evaluations of the Tabu Search. Every admitted candidate is still
scored exactly, so a misclassification can only cause a good candidate to be skipped, never an
incorrect fitness value. Once the tiling is fixed, a convex amplitude--phase optimization of the
domino excitations improves the directivity of the scanned beam under a sidelobe constraint.

The main contributions of this paper are as follows.
\begin{enumerate}
	\item A local-flip Markov chain generator, which converges to the uniform distribution over
	all domino tilings, produces balanced tilings (close to equal numbers of horizontal and
	vertical dominoes) about $5.2\times$ faster than a randomized greedy baseline.
	\item A Tabu Search engine, selected through a comparison with a genetic algorithm, simulated
	annealing, and iterated local search over eight independent seeds, reaches a mean SLL about
	$0.8$~dB lower than the other three at comparable directivity.
	\item A CNN surrogate trained on a two-channel structural encoding of the tilings gates the
	Tabu Search. For apertures from $20\times20$ to $100\times100$, it reduces the full
	evaluations by $60.6$--$70.6\%$ and the search time by $2.56$--$3.80\times$ with equal or
	slightly higher fitness, and a single run including dataset generation and training is already
	faster than the standalone search for apertures of $50\times50$ and larger.
	\item A convex amplitude--phase optimization of the fixed tilings gives equal or higher
	directivity than two state-of-the-art methods at a comparable sidelobe level, with more than
	an order of magnitude shorter computation time. It also compensates for a tiling that is not
	matched to the scan angle.
	\item A $36\times36$ dual-band circularly polarized array for low Earth orbit~(LEO) terminals
	at $11.2$ and $14.0$~GHz uses $648$ dominoes per band and a common element spacing selected
	from a sweep. Compared with a reference irregular array, it has no grating lobe up to a scan
	angle of $50^\circ$, higher directivity over most of the scan range, and a spacing that is
	more practical for the lower band.
\end{enumerate}

The remainder of this paper is organized as follows. Section~\ref{sec:formulation} presents the
array model, the tiling generator, and the pattern metrics. Section~\ref{sec:hybrid} describes
the optimization problem, the Tabu Search, the CNN surrogate, and the hybrid search.
Section~\ref{sec:numerical_results} reports the numerical results, and
Section~\ref{sec:conclusion} concludes the paper.

\section{Mathematical Formulation}
\label{sec:formulation}

\subsection{Array Model}
\label{sec:metrics}

Consider a planar array of $M \times N$ elements on a rectangular lattice in the $(x,y)$-plane
with spacing $d_x = d_y = d = \lambda/2$ unless stated otherwise, where $\lambda$ is the
free-space wavelength. Each element occupies a square unit cell of side $d$, referred to as a
\textit{pixel}. The array factor at the direction cosines
$(u,v) = (\sin\theta\cos\varphi,\,\sin\theta\sin\varphi)$ is
\begin{equation}
	F(u,v) = \sum_{m=1}^{M}\sum_{n=1}^{N}
	w_{mn}\,e^{-jk(m d\, u \,+\, n d\, v)},
	\label{eq:AF}
\end{equation}
where $k = 2\pi/\lambda$ and $w_{mn}=\alpha_{mn}e^{j\beta_{mn}}\in\mathbb{C}$ is the excitation
of the $(m,n)$-th element. To steer the main beam to $(\theta_0,\varphi_0)$, the phase is set to
$\beta_{mn}=k\,d\,(m u_0 + n v_0)$ with $u_0=\sin\theta_0\cos\varphi_0$ and
$v_0=\sin\theta_0\sin\varphi_0$ (steering phase). This phase is used in the tiling search;
in the excitation optimization of Section~\ref{sec:scalability} the phases are free.

The directivity and the sidelobe level (SLL) are computed from the normalized pattern
$\tilde{F}(u,v)=|F(u,v)|/\max|F|$ over the visible region
$\Omega=\{(u,v)\mid u^2+v^2\leq1\}$, with the SLL taken outside a circular exclusion disk
$\mathcal{D}_{\mathrm{ex}}$ around the main beam $(u_0,v_0)$.
\subsection{Domino Sub-Array Architecture}
\label{sec:domino_model}

To reduce the number of control channels, the $M\times N$ pixels ($MN$ even, so
that a complete domino partition exists) are grouped into $Q = MN/2$
non-overlapping domino tiles, each covering two 4-connected adjacent pixels and
driven by a single transmit/receive module~(TRM). The partition is encoded by
the clustering vector $\mathbf{c} \triangleq \{c_{mn};\, m=1,\ldots,M;\,
n=1,\ldots,N\}$, where $c_{mn}\in\{1,\ldots,Q\}$ identifies the tile to which
pixel $(m,n)$ belongs, subject to the full-coverage constraints
\begin{equation}
	\sum_{q=1}^{Q} \delta_{c_{mn},q} = 1
	\;\; \forall\,(m,n),
	\qquad
	\sum_{m=1}^{M}\sum_{n=1}^{N} \delta_{c_{mn},q} = 2
	\;\; \forall\, q,
	\label{eq:constraints}
\end{equation}
where $\delta_{c_{mn},q}$ is the Kronecker delta. Since each tile shares a
single TRM, its two pixels are assigned a common complex weight $w_q$, obtained
by averaging the steered weights of those pixels,
\begin{equation}
	w_{mn} = \sum_{q=1}^{Q} \delta_{c_{mn},q}\, w_q,
	\qquad
	w_q \triangleq
	\frac{w_{m_1^{(q)} n_1^{(q)}} + w_{m_2^{(q)} n_2^{(q)}}}{2},
	\label{eq:domino_avg}
\end{equation}
where $(m_1^{(q)},n_1^{(q)})$ and $(m_2^{(q)},n_2^{(q)})$ denote the two pixels
of the $q$-th tile. This reduces the $MN$-dimensional weight vector to a
$Q$-dimensional one, a two-fold reduction in TRM count. Substituting
\eqref{eq:domino_avg} into \eqref{eq:AF}, the array factor becomes
\begin{equation}
	F(u,v) = \sum_{q=1}^{Q} w_q\, \Gamma_q(u,v),
	\label{eq:AF_tile}
\end{equation}
where the $q$-th \textit{tile pattern} is
\begin{equation}
	\Gamma_q(u,v) \triangleq
	\sum_{\substack{(m,n):\\ \delta_{c_{mn},q}=1}}
	e^{-jk(m d_x\, u \,+\, n d_y\, v)}.
	\label{eq:tile_pat}
\end{equation}

The set of all clustering vectors satisfying \eqref{eq:constraints} for a given
$M\times N$ aperture is denoted $\mathcal{T}_{M,N}$, and an element
$\mathbf{c}\in\mathcal{T}_{M,N}$ is referred to as a \textit{domino tiling}.
This set should not be confused with the visible $(u,v)$ region $\Omega$
introduced in Section~\ref{sec:metrics}.

\subsection{Domino Tiling Algorithm}
\label{sec:tiling_algo}

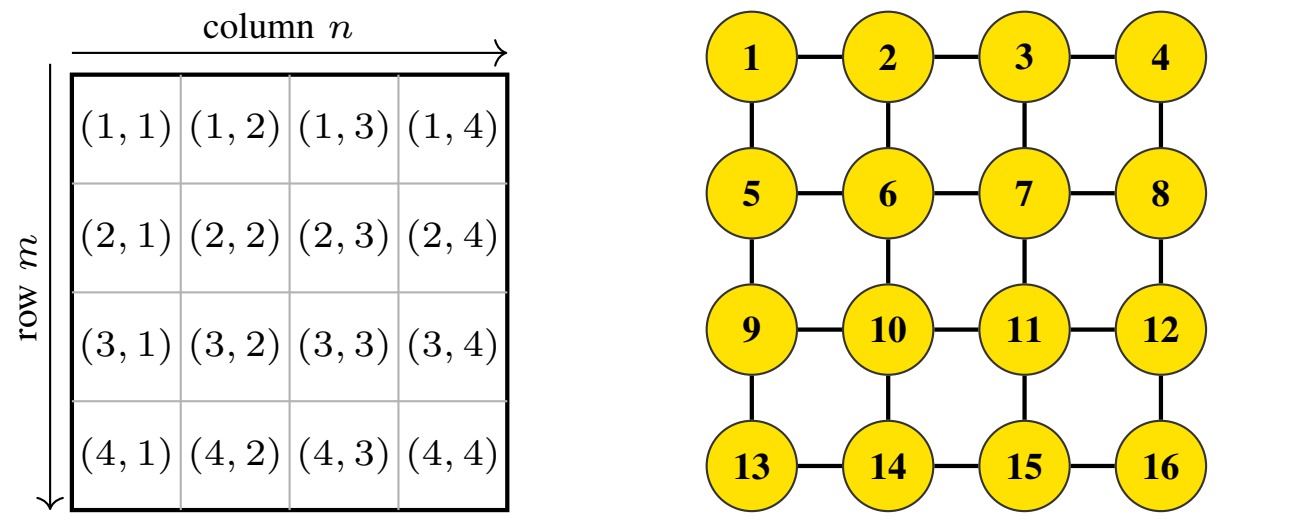
\begin{figure}[t]
	\centering
	\begin{subfigure}[b]{0.46\columnwidth}
		\centering
		\begin{tikzpicture}[scale=0.72]
			
			\draw[thick] (0,0) rectangle (4,4);
			\foreach \x in {1,2,3} \draw[thin,gray!60] (\x,0)--(\x,4);
			\foreach \y in {1,2,3} \draw[thin,gray!60] (0,\y)--(4,\y);
			
			\foreach \m/\ypos in {1/3.5, 2/2.5, 3/1.5, 4/0.5}
			\foreach \n/\xpos in {1/0.5, 2/1.5, 3/2.5, 4/3.5}
			\node[font=\scriptsize] at (\xpos,\ypos)
			{$(\m,\n)$};
			
			\draw[->] (-0.2,4.1) -- (-0.2,0);
			\node[font=\scriptsize, rotate=90] at (-0.4,2.05) {row $m$};
			
			\draw[->] (0,4.2) -- (4,4.2);
			\node[font=\scriptsize] at (1.9,4.45) {column $n$};
		\end{tikzpicture}
		\caption{Pixel grid $(m,n)$}
		\label{fig:pixel_grid}
	\end{subfigure}
	\hfill
	\begin{subfigure}[b]{0.50\columnwidth}
		\centering
		\begin{tikzpicture}[
			scale=0.82,
			vnode/.style={
				circle, draw=black!80, fill=yellow!85!orange,
				minimum size=17pt, inner sep=0pt,
				font=\scriptsize\bfseries
			}
			]
			
			\foreach \r/\c/\k in {%
				1/1/1,  1/2/2,  1/3/3,  1/4/4,%
				2/1/5,  2/2/6,  2/3/7,  2/4/8,%
				3/1/9,  3/2/10, 3/3/11, 3/4/12,%
				4/1/13, 4/2/14, 4/3/15, 4/4/16}{
				\pgfmathsetmacro\px{(\c-1)*1.1}
				\pgfmathsetmacro\py{(4-\r)*1.1}
				\node[vnode] (v\k) at (\px,\py) {\k};
			}
			
			\foreach \a/\b in {%
				1/2, 2/3, 3/4,
				5/6, 6/7, 7/8,
				9/10,10/11,11/12,
				13/14,14/15,15/16}
			\draw[thick] (v\a)--(v\b);
			
			\foreach \a/\b in {%
				1/5, 2/6,  3/7,  4/8,%
				5/9, 6/10, 7/11, 8/12,%
				9/13,10/14,11/15,12/16}
			\draw[thick] (v\a)--(v\b);
		\end{tikzpicture}
		\caption{Graph $\mathcal{G}_{4,4}=(\mathcal{V},\mathcal{E})$}
		\label{fig:grid_graph}
	\end{subfigure}
	\caption{Aperture modelling for a $4\!\times\!4$ array.
		(a)~Pixel coordinate grid $(m,n)$ with $M=N=4$.
		(b)~Corresponding grid graph $\mathcal{G}_{4,4}=(\mathcal{V},\mathcal{E})$
		with $|\mathcal{V}|=16$ nodes $v_{mn}$ and $|\mathcal{E}|=24$ edges
		encoding 4-connectivity.}
	\label{fig:aperture_model}
\end{figure}

Generating a clustering vector $\mathbf{c}$ that satisfies
\eqref{eq:constraints} is equivalent to producing a \textit{perfect matching} of
the $M\times N$ grid graph $\mathcal{G}_{M,N} = (\mathcal{V},\mathcal{E})$, whose
vertices are the pixels and whose edges connect 4-connected neighbors; each
matched edge defines one domino tile. Fig.~\ref{fig:aperture_model}
illustrates this correspondence for a $4\times 4$ aperture.

The generator must also avoid a systematic preference for one domino
orientation. Since the array factor is evaluated at several steering azimuths
$\varphi_0 \in \{0^\circ,45^\circ,90^\circ,135^\circ\}$, an orientation-biased
generator would couple asymmetrically with the steering phase gradient along $x$
and $y$, giving spuriously unequal sidelobe-level statistics at azimuths that
are equivalent by the symmetry of the square aperture (e.g., $\varphi_0=0^\circ$
and $\varphi_0=90^\circ$).

\subsubsection{Local-Flip Generation}

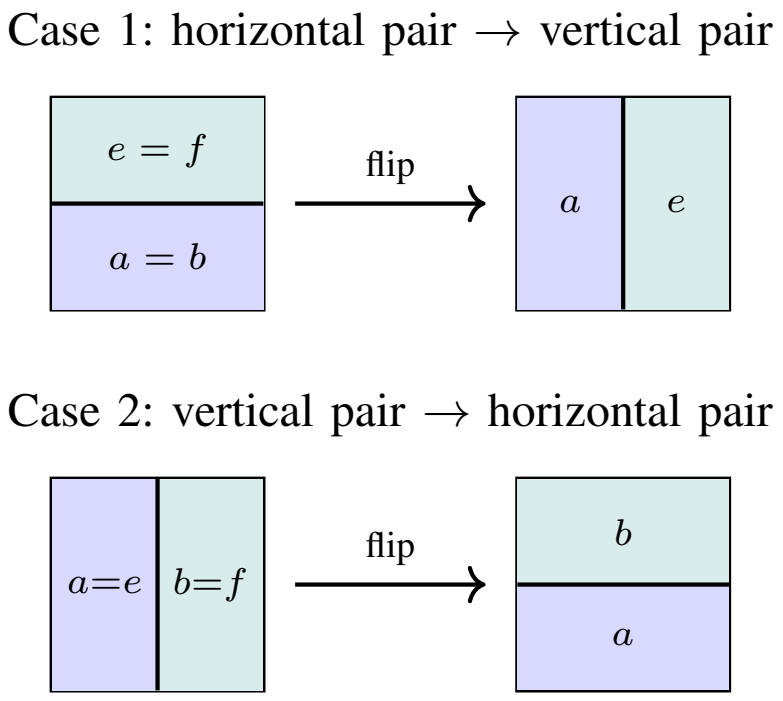
\begin{figure}[t]
	\centering
	\begin{tikzpicture}[
		scale=0.7,
		dom/.style={draw, thick, minimum width=1cm, minimum height=1cm,
			font=\scriptsize\bfseries, inner sep=0pt},
		lab/.style={font=\scriptsize}
		]
		
		\draw[thick] (0,0) rectangle (2,2);
		\fill[blue!15] (0,0) rectangle (2,1);
		\fill[teal!15] (0,1) rectangle (2,2);
		\draw[thick] (0,1)--(2,1);
		\node[lab] at (1,0.5) {$a=b$};
		\node[lab] at (1,1.5) {$e=f$};
		
		\draw[->,thick] (2.3,1) -- node[above,lab]{flip} (4.1,1);
		
		\draw[thick] (4.4,0) rectangle (6.4,2);
		\fill[blue!15] (4.4,0) rectangle (5.4,2);
		\fill[teal!15] (5.4,0) rectangle (6.4,2);
		\draw[thick] (5.4,0)--(5.4,2);
		\node[lab] at (4.9,1) {$a$};
		\node[lab] at (5.9,1) {$e$};
		\node[font=\small] at (3.2,2.6) {Case 1: horizontal pair $\to$ vertical pair};
		
		\begin{scope}[yshift=-3.6cm]
			
			\draw[thick] (0,0) rectangle (2,2);
			\fill[blue!15] (0,0) rectangle (1,2);
			\fill[teal!15] (1,0) rectangle (2,2);
			\draw[thick] (1,0)--(1,2);
			\node[lab] at (0.5,1) {$a{=}e$};
			\node[lab] at (1.5,1) {$b{=}f$};
			
			\draw[->,thick] (2.3,1) -- node[above,lab]{flip} (4.1,1);
			
			\draw[thick] (4.4,0) rectangle (6.4,2);
			\fill[blue!15] (4.4,0) rectangle (6.4,1);
			\fill[teal!15] (4.4,1) rectangle (6.4,2);
			\draw[thick] (4.4,1)--(6.4,1);
			\node[lab] at (5.4,0.5) {$a$};
			\node[lab] at (5.4,1.5) {$b$};
			\node[font=\small] at (3.2,2.6) {Case 2: vertical pair $\to$ horizontal pair};
		\end{scope}
		
	\end{tikzpicture}
	\caption{Illustration of the local-flip rule (Algorithm~\ref{alg:flip_tiling},
		Eq.~\eqref{eq:flip_rule}) on a $2\times2$ window with cell values
		$a=c_{m,n}$, $b=c_{m+1,n}$, $e=c_{m,n+1}$, $f=c_{m+1,n+1}$ (bottom-left,
		bottom-right, top-left, top-right, respectively). \textbf{Case 1:} a
		vertically-stacked pair of horizontal dominoes ($a=b$, $e=f$) is converted
		into a horizontally-adjacent pair of vertical dominoes. \textbf{Case 2:}
		the converse move. All other configurations are left unchanged.}
	\label{fig:flip_rule}
\end{figure}

We generate tilings with a local-move Markov chain acting on
$\mathcal{T}_{M,N}$, following the construction of Luby
\textit{et al.}~\cite{lubyRandallSinclair2001} for domino and rhombus tilings of
planar lattice regions. The chain is initialized at the deterministic
\textit{brick} tiling obtained by pairing pixels along the $m$-axis (assumed,
without loss of generality, to have even length),
\begin{equation}
	c_{m,n} = \left\lceil \frac{m}{2} \right\rceil + \frac{M}{2}(n-1),
	\qquad m=1,\ldots,M,\ n=1,\ldots,N,
	\label{eq:tiling_init}
\end{equation}
and is then evolved for $K$ iterations. At each iteration, a $2\times2$ window
with top-left pixel $(m,n)$ is proposed, with $m$ drawn uniformly from
$\{1,\ldots,M-1\}$ and $n$ from $\{1,\ldots,N-1\}$, and a \textit{flip} is
applied whenever the window is covered by two parallel, like-oriented dominoes.
Writing $a=c_{m,n}$, $b=c_{m+1,n}$, $e=c_{m,n+1}$, $f=c_{m+1,n+1}$ for the four
tile indices in the window, the flip rule is
\begin{equation}
	(c_{m+1,n},\,c_{m,n+1}) \leftarrow
	\begin{cases}
		(e,\,a), & a=b,\ e=f,\ a\neq e,\\[2pt]
		(a,\,b), & a=e,\ b=f,\ a\neq b,\\[2pt]
		(b,\,e), & \text{otherwise (no flip)}.
	\end{cases}
	\label{eq:flip_rule}
\end{equation}
The first case converts a vertically-stacked pair of horizontal dominoes into a
horizontally-adjacent pair of vertical dominoes, and the second case is its
converse, as illustrated in Fig.~\ref{fig:flip_rule}. The procedure is
summarized in Algorithm~\ref{alg:flip_tiling}.

\begin{algorithm}[t]
	\fontsize{8}{10}
	\caption{Local-Flip Domino Tiling Generator}
	\label{alg:flip_tiling}
	\begin{algorithmic}[1]
		\Require Aperture size $M \times N$ ($M$ even), flip budget $K$
		\Ensure  Tiling $\mathbf{c} = \{c_{mn}\}$, $c_{mn} \in \{1,\dots,Q\}$, $Q = MN/2$
		\State Initialize $\mathbf{c}$ via \eqref{eq:tiling_init}
		\For{$t = 1$ to $K$}
		\State Draw $m \sim \mathrm{Unif}\{1,\dots,M{-}1\}$,\;
		$n \sim \mathrm{Unif}\{1,\dots,N{-}1\}$
		\State $a \leftarrow c_{m,n}$,\; $b \leftarrow c_{m+1,n}$,\;
		$e \leftarrow c_{m,n+1}$,\; $f \leftarrow c_{m+1,n+1}$
		\If{$a = b$ and $e = f$ and $a \neq e$}
		\State $c_{m+1,n} \leftarrow e$;\quad $c_{m,n+1} \leftarrow a$
		\Statex \hspace{3.05em}\textcolor{gray!70!black}{$\triangleright$ horizontal pair $\to$ vertical pair}
		\ElsIf{$a = e$ and $b = f$ and $a \neq b$}
		\State $c_{m+1,n} \leftarrow a$;\quad $c_{m,n+1} \leftarrow b$
		\Statex \hspace{3.05em}\textcolor{gray!70!black}{$\triangleright$ vertical pair $\to$ horizontal pair}
		\Else
		\Statex \hspace{2.3em}\textcolor{gray!70!black}{$\triangleright$ window not flippable; leave $\mathbf{c}$ unchanged}
		\EndIf
		\EndFor
	\end{algorithmic}
\end{algorithm}

\subsubsection{Statistical Properties}

The state space $\mathcal{T}_{M,N}$ under the moves in \eqref{eq:flip_rule} is connected for
simply-connected planar regions such as the rectangle considered here: any two tilings can be
joined by a finite sequence of local flips, a consequence of Thurston's height-function
representation of domino tilings~\cite{thurston1990conway}. Since each flip is applied to a
window drawn uniformly at random and is deterministic given the window, the chain is reversible
with respect to the uniform measure on $\mathcal{T}_{M,N}$, which is therefore its unique
stationary distribution~\cite{lubyRandallSinclair2001}. Polynomial bounds on the number of flips
needed for convergence are known for closely related dimer
models~\cite{wilson2004mixing, laslierToninelli2015}; in practice, we use $20MN$ flip proposals
per tiling.

\begin{figure}[!t]
	\centering
	\includegraphics[width=0.75\linewidth]{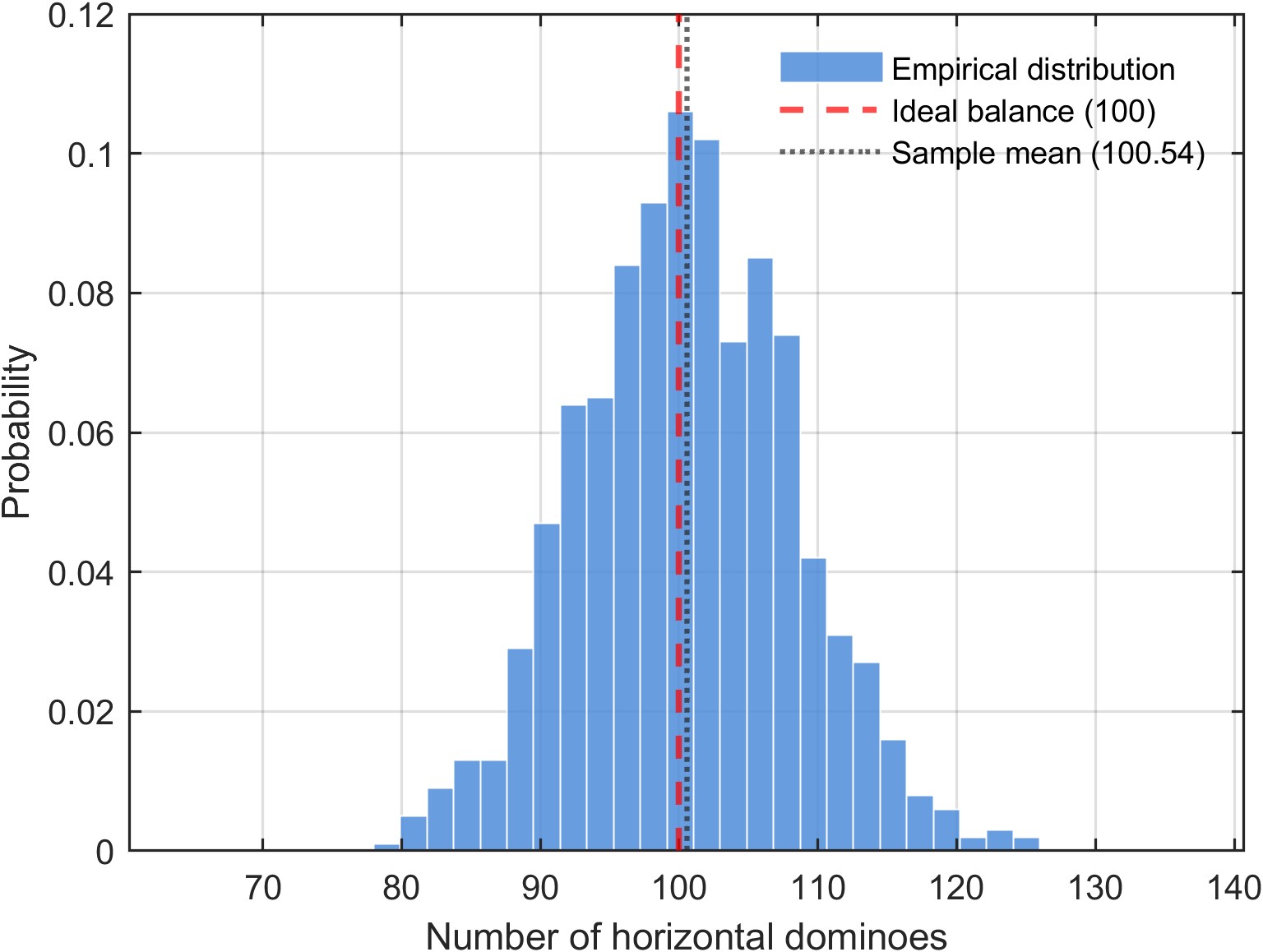}
	\caption{Distribution of the number of horizontal dominoes over $1000$ tilings generated by
		the local-flip generator (Algorithm~\ref{alg:flip_tiling}), $M=N=50$. The empirical mean
		closely matches the balanced value ($625$).}
	\label{fig:orientation_balance}
\end{figure}

Fig.~\ref{fig:orientation_balance} is consistent with this: over $1000$ generated tilings of a
$50\times50$ aperture, the number of horizontal dominoes is tightly concentrated around the
balanced value $Q/2=625$, with no systematic orientation bias.

We also compare the local-flip generator with a randomized greedy matching baseline, which visits
pixels in raster order, pairs each unmatched pixel with a random free 4-neighbor, and restarts
whenever isolated pixels remain. Both were implemented in MATLAB and run on the same machine for a
$50\times50$ aperture with $10^4$ tilings (Table~\ref{tab:generator_comparison}).

\begin{table}[!t]
	\centering
	\caption{Tiling Generator Comparison for a $50\times50$ Aperture}
	\label{tab:generator_comparison}
	\footnotesize
	\renewcommand{\arraystretch}{1.15}
	\begin{threeparttable}
		\begin{tabular}{lcc}
			\hline\hline
			& \textbf{Local-flip} & \textbf{Greedy + restart} \\
			\hline
			Total time (s)              & $403$  & $\approx 2100$ \\
			Time per tiling (ms)        & $\approx 40$ & $\approx 210$ \\
			Relative speed              & $5.2\times$ & $1\times$ \\
			Valid tiling guaranteed     & Always & After restarts \\
			Sampling distribution       & Uniform (in the limit) & Not guaranteed \\
			\hline\hline
		\end{tabular}
		\begin{tablenotes}\footnotesize
			\item Each generator produced $10^4$ samples.
		\end{tablenotes}
	\end{threeparttable}
\end{table}

The local-flip generator is about $5.2\times$ faster, because it starts from a
valid tiling and only applies local moves, so no pass is ever discarded. The
greedy baseline wastes most of its effort on failed passes, and its fixed
visiting order gives early pixels priority and offers no uniformity guarantee,
whereas the local-flip chain converges to the uniform distribution over
$\mathcal{T}_{M,N}$. Since the CNN surrogate requires a large labeled dataset,
this speed-up directly reduces the cost of dataset generation.

\section{Proposed Hybrid Optimization Framework}
\label{sec:hybrid}

\subsection{Optimization Problem Formulation}
\label{sec:opt_problem_formulation}

The design variable is a domino tiling $\mathcal{T}\in\mathcal{T}_{M,N}$ of the
$M\times N$ aperture into $MN/2$ non-overlapping dominoes (Section~\ref{sec:domino_model}).
Each tiling induces a locally amplitude-perturbed excitation obtained by pairwise
averaging the Chebyshev-tapered weights within each domino. Since every tiling is
reachable from any other through the local-flip operator, which preserves validity,
the optimization is unconstrained over $\mathcal{T}_{M,N}$.

To support beam steering, the array must keep an acceptable pattern simultaneously at $P$
directions $(\theta_0,\varphi_p)$, $p=1,\dots,P$ (here $\theta_0=45^\circ$ and
$\varphi_p\in\{0^\circ,45^\circ,90^\circ,135^\circ\}$). For a candidate tiling $\mathcal{T}$, the
directivity $D_p(\mathcal{T})$ and peak sidelobe level $\mathrm{SLL}_p(\mathcal{T})$ (in dB relative
to the main-beam peak, negative-valued) at each direction are computed from the array-factor model
of Section~\ref{sec:formulation} and combined into a single fitness
\begin{equation}
	F(\mathcal{T}) = \bar{D}(\mathcal{T}) - \overline{\mathrm{SLL}}(\mathcal{T}),
	\label{eq:fitness}
\end{equation}
where $\bar{D}$ and $\overline{\mathrm{SLL}}$ are the means over the $P$ directions. Maximizing $F$
raises the directivity and lowers the sidelobes, and the design task is
$\mathcal{T}^\star=\arg\max_{\mathcal{T}}F(\mathcal{T})$. Only these two metrics are used because
the comparisons in this work mainly concern them; other metrics, such as the beamwidth or the
average sidelobe level, can be added to $F$ in the same way. The scalarized form gives a single
ranking criterion for the statistical comparison of the metaheuristics. The search space is
discrete and non-convex, and each evaluation of $F$ requires a full multi-angle array-factor
computation, which dominates the cost of the optimization loop.

\subsection{Tabu Search Optimization Engine}
\label{sec:tabu_search}

Tabu Search~\cite{glover1989tabu1} is adopted as the search engine. It operates directly on
$\mathcal{T}$ using the local-flip neighborhood, with a memory-guided single trajectory.
A $2\times2$ block at position $(r,j)$ is \emph{flippable} if it holds two parallel dominoes;
flipping rotates the pair and leaves the rest of the tiling unchanged. The neighborhood
$\mathcal{N}(\mathcal{T})$ is the set of all flippable blocks. Because it is large for big
apertures, exhaustive scoring is impractical, so at every iteration $n_s$ blocks are sampled
uniformly from $\mathcal{N}(\mathcal{T})$ and scored with \eqref{eq:fitness}. The value of
$n_s$ is chosen so that the per-step evaluation cost matches that of the other benchmarked
metaheuristics.

To avoid cycling, a block $(r,j)$ that has just been flipped is \emph{tabu} for $\tau$
iterations (the tabu tenure), i.e., its reverse move is forbidden. The best sampled candidate
is accepted if it is non-tabu, or if it is tabu but better than the best fitness found so
far (aspiration criterion). If neither holds for any sampled candidate, a single random flip
is applied for diversification, without a fitness evaluation.

Algorithm~\ref{alg:tabu} summarizes the procedure. It tracks the current tiling
$\mathcal{T}_c$, which may accept non-improving moves, and the incumbent best tiling
$\mathcal{T}^\star$, which never regresses. The counter $\mathrm{used}$ is the number of
full evaluations performed, and the search stops when it reaches the budget $B$. Since every
sampled candidate is evaluated, $B$ here bounds both the generated candidates and the full
evaluations. This no longer holds for the CNN-gated search of Section~\ref{sec:hybrid_framework}.

\begin{algorithm}[t]
	\fontsize{8}{10}
	\caption{Tabu Search for Domino Tiling Optimization}
	\label{alg:tabu}
	\begin{algorithmic}[1]
		\Require Initial tiling $\mathcal{T}_c$, tabu tenure $\tau$, neighborhood sample size
		$n_s$, evaluation budget $B$
		\Ensure  Best tiling found $\mathcal{T}^\star$
		\State $\mathcal{T}^\star \leftarrow \mathcal{T}_c$;\quad tabu list $\leftarrow \emptyset$;\quad $\mathrm{used} \leftarrow 0$
		\While{$\mathrm{used} < B$}
		\State Sample $n_s$ flippable blocks $\{(r_i,j_i)\}_{i=1}^{n_s}$ from $\mathcal{N}(\mathcal{T}_c)$
		\For{$i = 1$ to $n_s$}
		\State $\mathcal{T}_i' \leftarrow \mathrm{flip}(\mathcal{T}_c, r_i, j_i)$;\quad evaluate $F(\mathcal{T}_i')$;\quad $\mathrm{used} \leftarrow \mathrm{used}+1$
		\EndFor
		\State $\mathcal{T}' \leftarrow \arg\max_i F(\mathcal{T}_i')$ over candidates that are non-tabu or satisfy $F(\mathcal{T}_i')>F(\mathcal{T}^\star)$
		\If{such a candidate exists}
		\State $\mathcal{T}_c \leftarrow \mathcal{T}'$;\quad mark the reverse move tabu for $\tau$ iterations
		\If{$F(\mathcal{T}_c) > F(\mathcal{T}^\star)$}
		\State $\mathcal{T}^\star \leftarrow \mathcal{T}_c$
		\EndIf
		\Else
		\State Apply one random flip to $\mathcal{T}_c$ \Comment{diversification, no evaluation}
		\EndIf
		\EndWhile
	\end{algorithmic}
\end{algorithm}

\subsection{Metaheuristic Benchmarking and Optimizer Selection}
\label{sec:optimizer_benchmark}

To validate Tabu Search, four metaheuristics---Genetic Algorithm (GA), Simulated Annealing
(SA), Iterated Local Search (ILS), and Tabu Search---were benchmarked on the fitness
\eqref{eq:fitness} for a $20\times20$ aperture, with a matched budget of 2500 fitness
evaluations per run and eight independent seeds. All four use the same local-flip operator
and differ only in how candidate moves are proposed and accepted.

\begin{figure}[!t]
	\centering
	\includegraphics[width=1\columnwidth]{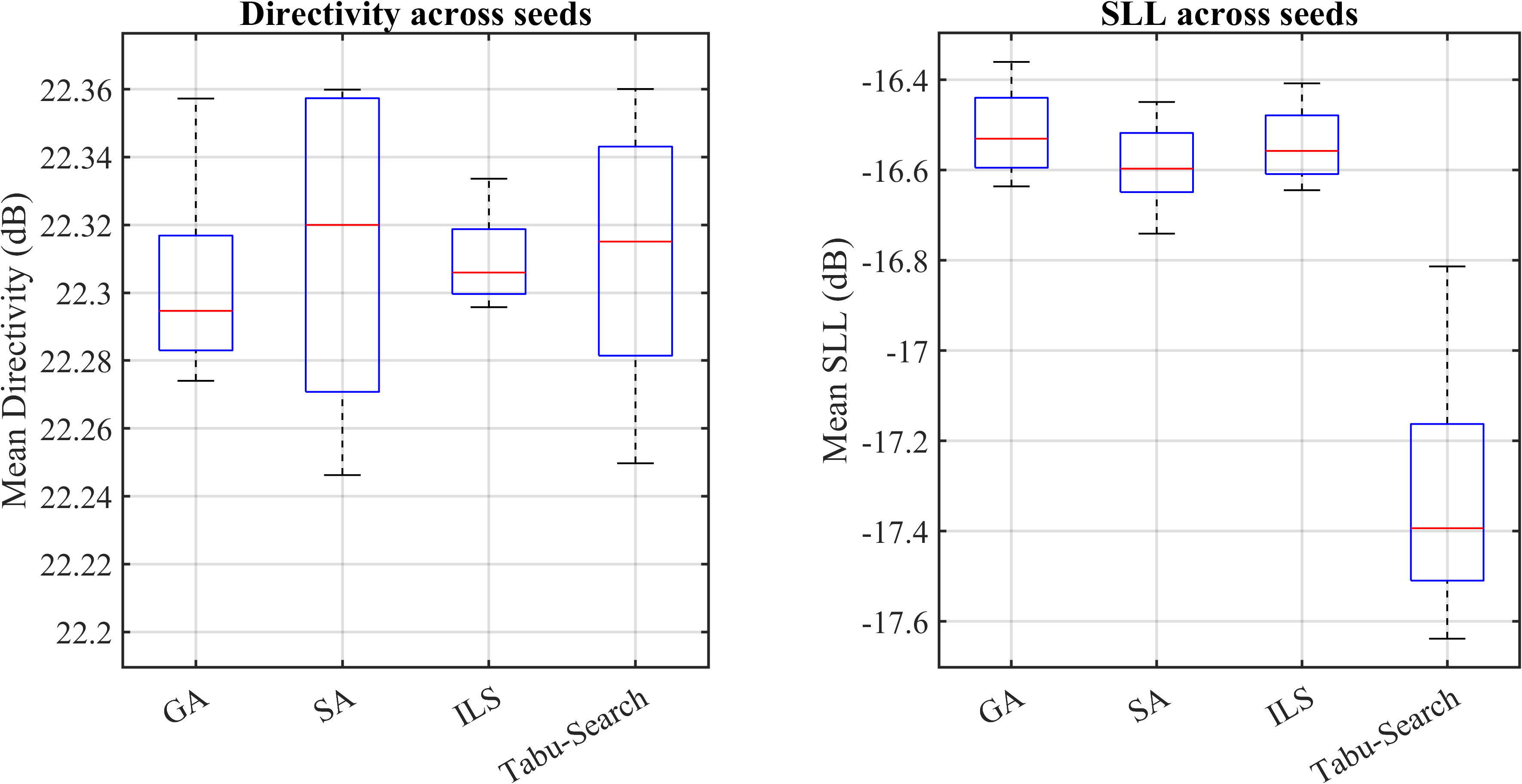}
	\caption{Optimizer comparison across eight independent seeds on a $20\times20$ domino-tiled
		aperture. Left: mean directivity. Right: mean sidelobe level.}
	\label{fig:opt_result}
\end{figure}

As shown in Fig.~\ref{fig:opt_result}, Tabu Search achieved the highest median fitness ($39.73$ vs.\ $38.81$, $38.93$, and $38.85$ for GA, SA, and ILS), mainly through lower sidelobes ($-17.4$~dB median vs.\ about $-16.5$~dB for the others) at comparable directivity. The difference is statistically significant against each of the other methods ($p<0.001$, Kruskal--Wallis and pairwise Mann--Whitney U tests with Holm--Bonferroni correction). This is likely because Tabu Search evaluates several candidate moves per step and picks the best allowed one, which helps it escape local optima within the same budget, whereas SA and ILS try a single candidate at a time.

\subsection{Deep Learning Surrogate Model}
\label{sec:cnn_surrogate}

\begin{figure*}[t]
	\centering
	\includegraphics[width=0.7\linewidth]{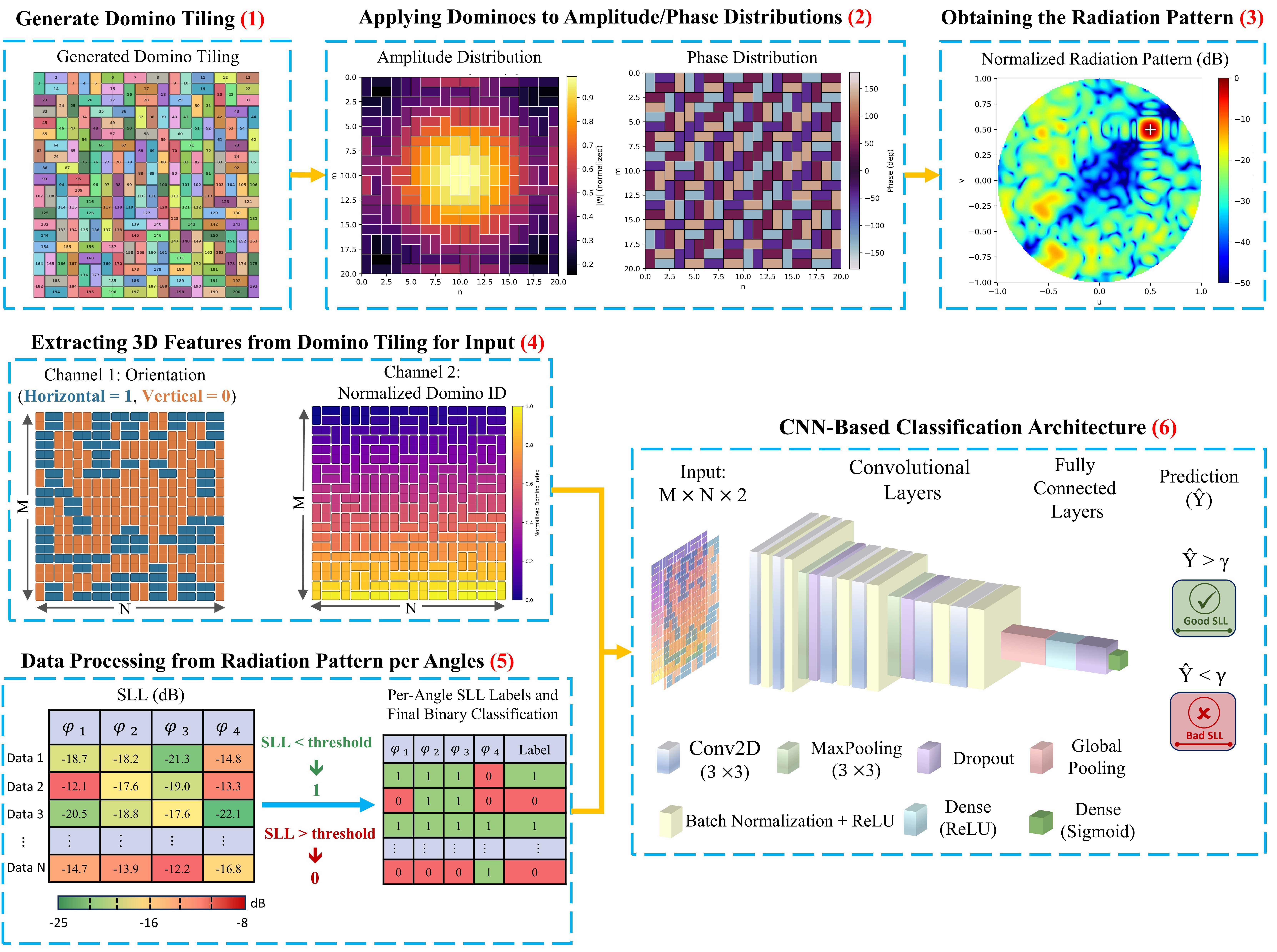}
	\caption{CNN surrogate pipeline: (1) tiling generation; (2) mapping onto Chebyshev-tapered
		excitations via domino averaging; (3) multi-angle radiation pattern; (4) two-channel
		feature extraction; (5) per-angle SLL thresholding and majority-vote labeling; (6) CNN
		producing a probability $\hat{Y}$, compared with the threshold $\gamma$ to classify the
		tiling as good or bad SLL.}
	\label{fig:cnn_pipeline}
\end{figure*}

Across tilings, the mean directivity varies only marginally (about $0.1$--$0.15$~dB for all
four optimizers), whereas the mean sidelobe level varies by more than $1$~dB. For a fixed
aperture and taper, the SLL is therefore the dominant factor separating a good tiling from a
poor one. Since surrogate models are widely used in antenna design to replace costly
evaluations~\cite{koziel2014expedited, easum2018efficient}, we develop a convolutional neural
network (CNN) that classifies a tiling, directly from its structure and without any
array-factor computation, as having a favorable multi-angle SLL or not
(Fig.~\ref{fig:cnn_pipeline}).

Each tiling is encoded as a two-channel $M\times N$ image. The first channel stores the
orientation of the domino covering each cell (horizontal $=1$, vertical $=0$), and the second
stores the normalized domino identifier $c_{mn}/Q$, with $Q=MN/2$, which gives a continuous cue
for domino boundaries and adjacency. The network needs neither the excitations nor the
radiation pattern as input.

Labels come from an array-factor evaluation at $P=4$ azimuth angles. At each angle, a tiling is
labeled favorable if its SLL is below a threshold set close to the median SLL of the dataset,
which balances the two classes, and the overall label is a majority vote (at least three of the
four angles), so that a favorable tiling must be robust across most, but not necessarily all,
steering directions.

The network outputs a scalar $\hat{Y}\in[0,1]$, the probability that the tiling is favorable.
It has three convolutional blocks with $16$, $32$, and $64$ filters, each with two $3\times3$
convolutional layers (L2 regularization, batch normalization, ReLU), $2\times2$ max-pooling, and
spatial dropout in the first two blocks. Global average pooling feeds a fully connected layer
($32$ units, dropout $0.4$) and a sigmoid output. Training uses the Adam optimizer with binary
cross-entropy, early stopping, and learning-rate reduction on the validation loss. At inference,
a tiling is classified as favorable if $\hat{Y}\geq\gamma$. A larger $\gamma$ gives a more
selective gate with fewer false positives, and a smaller $\gamma$ retains more candidates at the
cost of more full evaluations.

\subsection{Hybrid Deep Learning--Optimization Framework}
\label{sec:hybrid_framework}

The CNN acts as a cheap admissibility gate placed ahead of the expensive evaluation of $F$
inside the Tabu Search. To keep the comparison budget-fair, $B$ is redefined for the hybrid as
the number of candidate tilings \emph{generated} from the sampled neighborhoods, whether or not
they are evaluated. The number of full evaluations is tracked separately as
$n_{\mathrm{full}}\leq B$, with equality only if the gate admits every candidate. The hybrid
and ungated searches thus generate the same number of candidates and differ in how many are
fully evaluated.

\begin{figure}[!t]
	\centering
	\begin{tikzpicture}[
		node distance=0.25cm,
		font=\scriptsize,
		process/.style={rectangle, draw, rounded corners, fill=blue!8,
			text width=3.4cm, align=center, minimum height=0.5cm, inner sep=2pt},
		loopbox/.style={rectangle, draw, rounded corners, fill=green!10,
			text width=3.4cm, align=left, minimum height=1.2cm, inner sep=3pt},
		expensive/.style={rectangle, draw, rounded corners, fill=red!12,
			text width=3.4cm, align=center, minimum height=0.5cm, inner sep=2pt},
		decision/.style={diamond, draw, fill=yellow!15, aspect=3.2,
			text width=1.8cm, align=center, inner sep=0pt},
		term/.style={rectangle, draw, rounded corners, fill=blue!8,
			align=center, minimum height=0.45cm, inner sep=2pt},
		arr/.style={-{Latex[length=1.6mm]}, thick}
		]
		
		\node[process] (sample) {Sample $n_s$ candidates from local-flip neighborhood};
		
		\node[loopbox, below=of sample] (loop)
		{For each candidate $\mathcal{T}'$: generate ($B\mathrel{+}=1$), CNN predicts $\hat{Y}$;
			if $\hat{Y}>\gamma$, evaluate $F(\mathcal{T}')$ ($n_{\mathrm{full}}\mathrel{+}=1$) and
			track best non-tabu; else discard};
		
		\node[decision, below=of loop] (admitted) {Any admitted?};
		
		\node[expensive, below=of admitted] (fallback)
		{Random candidate, evaluated directly (gate bypassed)};
		
		\node[process, below=of fallback] (move)
		{Apply best move; mark tabu; update $\mathcal{T}^\star$};
		
		\node[decision, below=of move] (budget) {$B$ exhausted?};
		
		\node[term, below=of budget] (out) {Output $\mathcal{T}^\star$};
		
		\draw[arr] (sample) -- (loop);
		\draw[arr] (loop) -- (admitted);
		\draw[arr] (admitted) -- node[right]{no} (fallback);
		\draw[arr] (fallback) -- (move);
		\draw[arr] (admitted.east) -- node[above]{yes} ++(0.7cm,0) |- (move.east);
		\draw[arr] (move) -- (budget);
		\draw[arr] (budget) -- node[right]{yes} (out);
		\draw[arr] (budget.west) -- ++(-1.1cm,0) |- node[above,pos=0.3]{no} (sample.west);
		
	\end{tikzpicture}
	\caption{CNN-gated hybrid Tabu Search loop. Green: cheap CNN inference. Red: full
		array-factor evaluation, incurred only for admitted candidates or the fallback
		candidate. $B$ counts every generated candidate; $n_{\mathrm{full}}$ is the number of
		full evaluations.}
	\label{fig:hybrid_flowchart}
\end{figure}
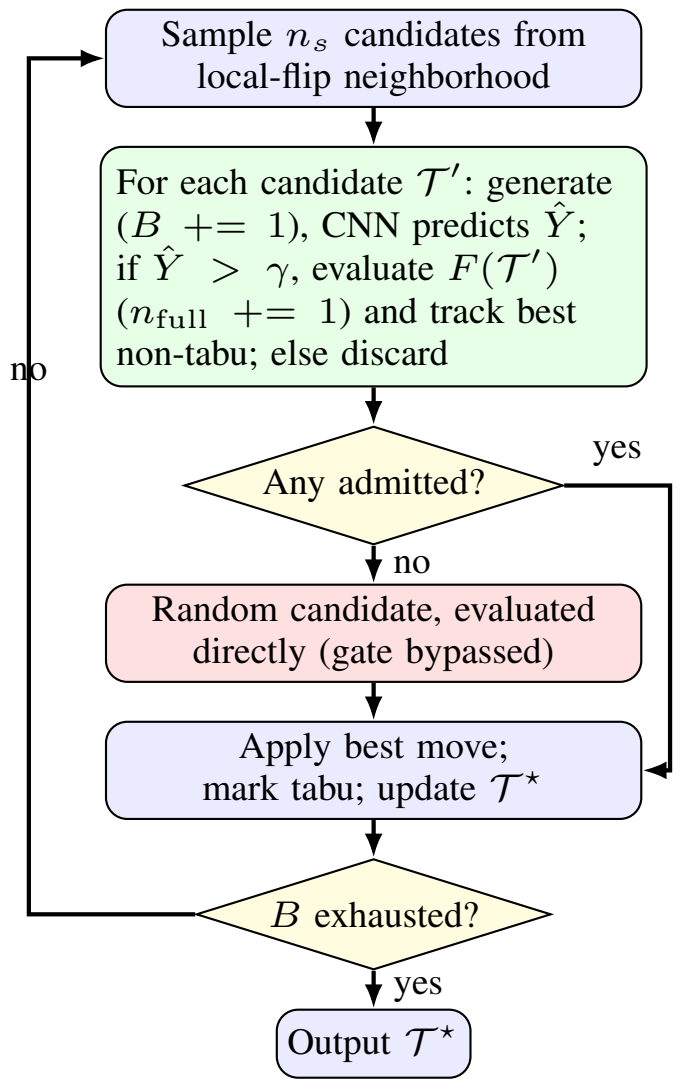

At each iteration, $n_s$ candidates are sampled from the neighborhood of the current tiling
and counted against $B$. The CNN predicts $\hat{Y}(\mathcal{T}')$ for each from the encoding
alone. Candidates with $\hat{Y}\geq\gamma$ are admitted and evaluated through $F$, and the
best admitted non-tabu candidate is tracked, with the aspiration criterion still applied.
The others are discarded without a full evaluation. If no candidate is admitted, one is drawn
uniformly from the batch and evaluated directly, so every iteration yields at least one true
evaluation and one accepted move. The loop is shown in Fig.~\ref{fig:hybrid_flowchart}, and the
resulting $n_{\mathrm{full}}$ is reported in Section~\ref{sec:numerical_results}.

\section{Numerical Results}
\label{sec:numerical_results}

This section evaluates the proposed framework in four stages. First,
Section~\ref{sec:surrogate_performance} reports the performance of the CNN surrogate. Second,
Section~\ref{sec:hybrid_vs_tabu} compares the CNN-gated hybrid Tabu Search with the standalone
Tabu Search for apertures of $20\times20$ to $100\times100$ elements, under the same random
seed and candidate budget $B$. Third, Section~\ref{sec:scalability} applies a convex
amplitude--phase optimization on top of the tilings found by the hybrid search and compares the
resulting arrays with two state-of-the-art tiling methods for large apertures. Finally,
Section~\ref{subsec:leo_application} applies the framework to a dual-band circularly polarized
$36\times36$ array for LEO satellite terminals. Unless stated otherwise, the first three stages
use the operating frequency and steering configuration of
Section~\ref{sec:opt_problem_formulation} ($\theta_0=45^\circ$,
$\varphi_p\in\{0^\circ,45^\circ,90^\circ,135^\circ\}$). The array-factor evaluations and both
Tabu Search variants were run in MATLAB R2023a on a 13th Gen Intel\textsuperscript{\textregistered}
Core\texttrademark{} i5-13400 CPU with 16~GB RAM, and the CNN surrogate was trained on a Google
Colaboratory GPU.

\subsection{Deep Learning Surrogate Performance}
\label{sec:surrogate_performance}

For each aperture ($20\times20$, $50\times50$, $80\times80$, and $100\times100$), an
independent dataset of $10{,}000$ tilings was generated with the local-flip procedure and
labeled as described in Section~III-D. The labeling threshold is the median SLL pooled over the
dataset and the four steering angles, so it is recomputed for each aperture and keeps the classes
balanced. Each dataset was split into stratified training, validation, and test subsets
($70\%/15\%/15\%$), and a separate CNN was trained for each aperture with early stopping on the
validation loss.

\begin{figure}[t]
	\centering
	\includegraphics[width=0.7\linewidth]{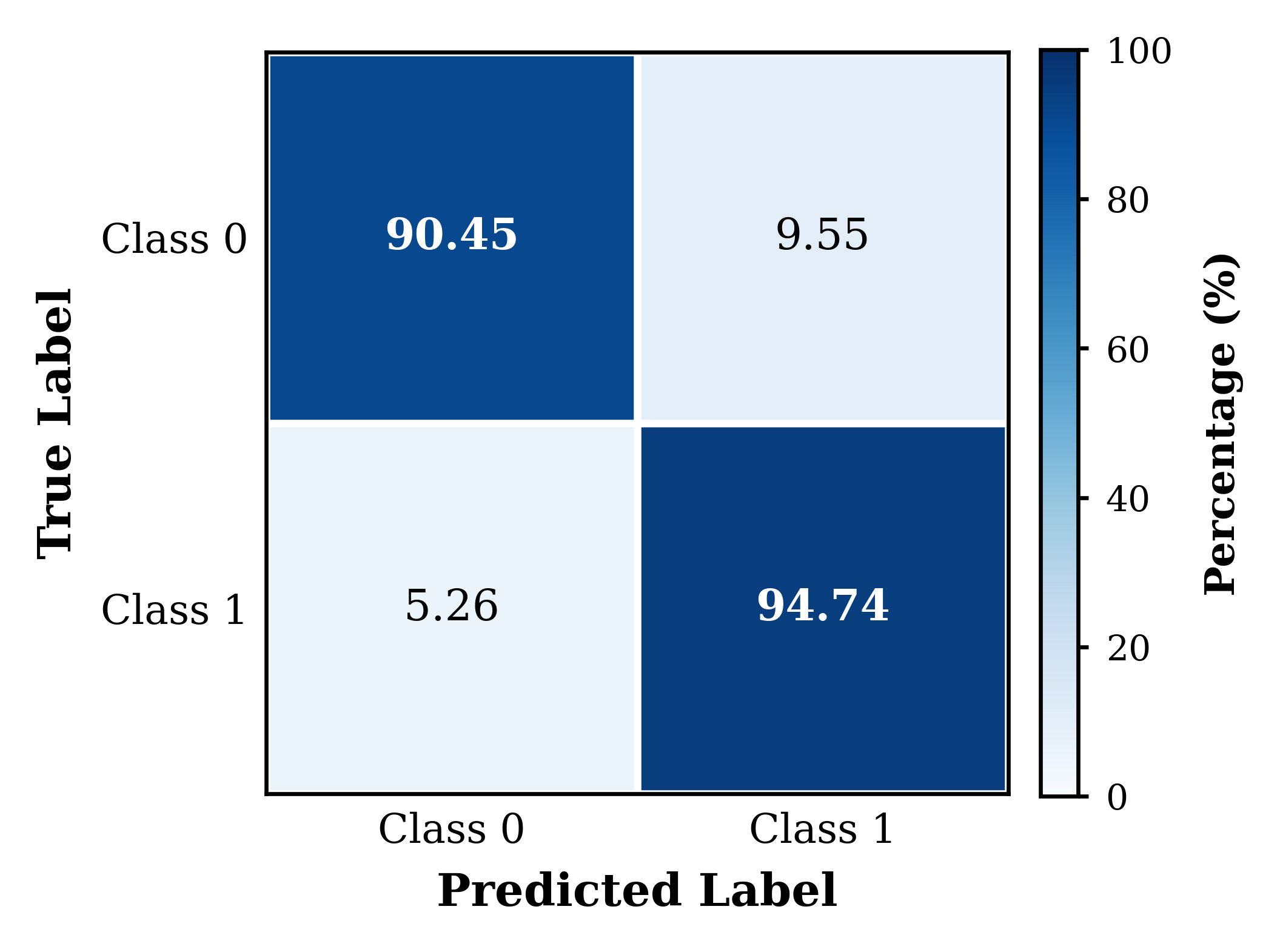}
	\caption{Row-normalized confusion matrix of the $50\times50$ CNN surrogate on the test set
		(Class 1: favorable SLL, Class 0: unfavorable SLL).}
	\label{fig:confusion_matrix}
\end{figure}

Table~\ref{tab:train_val_accuracy} reports the accuracy. The $50\times50$ surrogate is the most
accurate ($92.47\%$ validation) and the $20\times20$ surrogate the least ($82.62\%$). Accuracy is
not monotonic in aperture size because the taper and the threshold differ between apertures. The
gap between training and validation accuracy is below $1.9$ percentage points in all cases, so
there is no substantial overfitting. Fig.~\ref{fig:confusion_matrix} shows the row-normalized
confusion matrix of the $50\times50$ surrogate on the test set: $94.74\%$ of the favorable and
$90.45\%$ of the unfavorable tilings are classified correctly.

The surrogate need not be perfect to serve as a gate. Every admitted candidate is still scored
exactly by the array-factor model, so a misclassification can only cause a good candidate to be
skipped, never an incorrect fitness value.

\begin{table}[t]
	\centering
	\caption{Training and Validation Accuracy by Aperture Size}
	\label{tab:train_val_accuracy}
	\small
	\renewcommand{\arraystretch}{1.2}
	\begin{tabular}{lcc}
		\hline\hline
		\textbf{Aperture} & \textbf{Train Acc. (\%)} & \textbf{Val. Acc. (\%)} \\
		\hline
		$20\times20$   & $83.29$ & $82.62$ \\
		$50\times50$   & $92.90$ & $92.47$ \\
		$80\times80$   & $85.82$ & $87.70$ \\
		$100\times100$ & $85.07$ & $85.82$ \\
		\hline\hline
	\end{tabular}
\end{table}

\subsection{Hybrid CNN-Gated vs.\ Standalone Tabu Search}
\label{sec:hybrid_vs_tabu}

The array uses isotropic elements on a square lattice with $d=\lambda/2$. Each
tiling is evaluated for a beam scanned to $\theta_0=45^\circ$ at
$\varphi_p\in\{0^\circ,45^\circ,90^\circ,135^\circ\}$, and the fitness (8) is
averaged over these angles. The SLL is the highest sidelobe outside a circular
exclusion region of radius $0.15$ in the $(u,v)$ plane. Both searches use the
same Chebyshev taper ($-30$~dB for $20\times20$ to $80\times80$, $-35$~dB for
$100\times100$), candidate budget $B=8{,}000$, and random seed. For
$50\times50$, $\tau=15$, $n_s=30$, and $\gamma=0.97$. The $(u,v)$ grid is
refined with aperture size ($121$ to $380$ points per axis) so that the sidelobe
peak is not missed.

Table~\ref{tab:hybrid_vs_tabu_all} summarizes all four apertures. With the same candidate budget of $B=8{,}000$, the CNN gate reduces the number of full evaluations by $61.8$--$70.6\%$ (from $8{,}000$ to between $2{,}351$ and $3{,}156$) and the search time by $2.56$--$3.80\times$, while maintaining comparable final fitness, with a maximum difference of $0.14$ relative to the standalone search. The total time of the hybrid framework, including dataset generation, CNN training, and search, is $245$, $825$, $3{,}235$, and $6{,}818$~s for the $20\times20$, $50\times50$, $80\times80$, and $100\times100$ apertures, respectively.

\begin{table}[t]
	\centering
	\caption{Hybrid Versus Standalone Tabu Search}
	\label{tab:hybrid_vs_tabu_all}
	\small
	\renewcommand{\arraystretch}{1.2}
	\begin{tabular}{lccc}
		\hline\hline
		\textbf{Aperture} & \textbf{Full AF Evals.} & \textbf{Speed-up} & \textbf{Fitness Diff.} \\
		\hline
		$20\times20$   & $8000\to3059$ & $2.56\times$ & $+0.14$ \\
		$50\times50$   & $8000\to2351$ & $3.80\times$ & $0.00$ \\
		$80\times80$   & $8000\to2793$ & $2.75\times$ & $+0.03$ \\
		$100\times100$ & $8000\to3156$ & $2.56\times$ & $+0.04$ \\
		\hline\hline
	\end{tabular}
\end{table}

\begin{figure}[!t]
	\centering
	\includegraphics[width=0.65\linewidth]{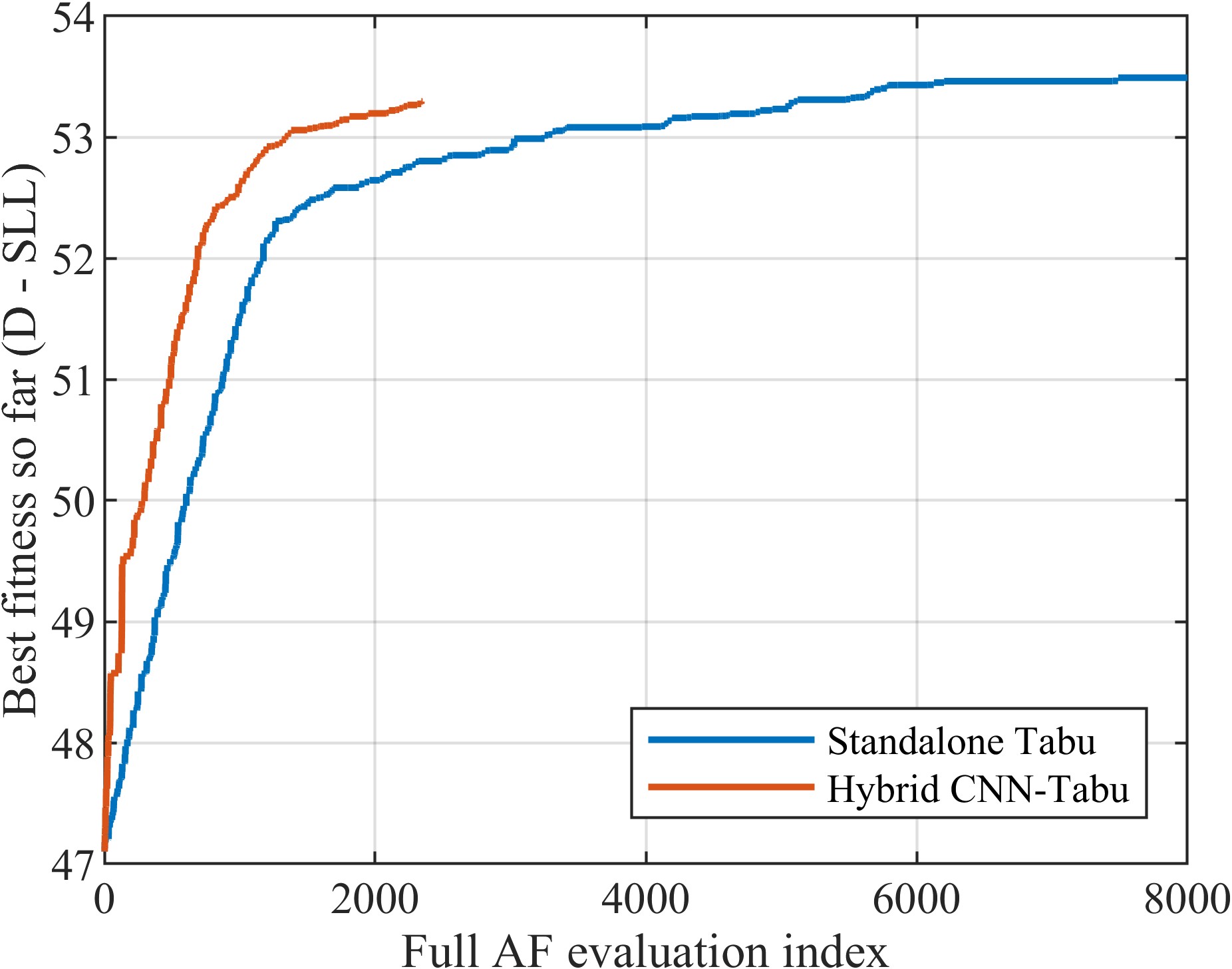}
	\caption{Best fitness so far versus the number of full AF evaluations for the
		standalone Tabu Search and the hybrid CNN--Tabu search ($50\times50$).}
	\label{fig:hybrid_convergence_50}
\end{figure}

Fig.~\ref{fig:hybrid_convergence_50} shows the best fitness found versus the
number of full array-factor (AF) evaluations, which dominate the cost, for
$50\times50$. At equal evaluation counts the hybrid is ahead. It reaches a
fitness of about $53.3$ (directivity $30.15$~dBi, SLL $-23.16$~dB) after
$2{,}351$ full evaluations, whereas the standalone search needs about $5{,}000$
to reach the same value.

\subsection{Larger-Scale Apertures}
\label{sec:scalability}

To assess the gain from also optimizing the excitation, a convex amplitude--phase
optimization was applied to the fixed final tilings of the $50\times50$, $80\times80$, and
$100\times100$ apertures, and the results were compared with two state-of-the-art tiling methods
for large arrays \cite{yang2024tiling,Anselmi2026}.

Each domino is fed by a single complex weight shared by its two elements. For each scan
direction, the weights maximize directivity subject to unit gain at the beam point, a sidelobe
bound over the visible region outside the main beam, and amplitude limits (a cap and a dynamic
range). The problem is a convex QCQP, except for the dynamic-range floor, which is enforced
heuristically. It was solved in its dual form by dual ascent, and the phases are not restricted
to the steering phase.

The sidelobe target was $-20.2$~dB for $50\times50$ and $-13.3$~dB for $80\times80$ and
$100\times100$, with a $0.25$~dB safety margin. The amplitude cap was $1.5$ times the peak
amplitude of the Chebyshev baseline, and the dynamic range was limited to $40$~dB. The
constraints were enforced on $241\times241$, $600\times600$, and $800\times800$ grids in the
$(u,v)$ plane and verified on finer grids of $400\times400$, $800\times800$, and
$1000\times1000$ points. Each scan angle took about $15$, $115$, and $230$~s, respectively. The
times in Table~\ref{tab:summary} are the total hybrid time (dataset generation, training, and
search) plus one convex optimization.

\begin{table*}[!t]
	\centering
	\caption{Summary of the Comparison With Previous Tiling Methods}
	\label{tab:summary}
	\renewcommand{\arraystretch}{1.2}
	\small
	\setlength{\tabcolsep}{4.5pt}
	\begin{threeparttable}
		\begin{tabular}{llccclcccccc}
			\hline\hline
			& & & & & & Normal & \multicolumn{4}{c}{Azimuth angle $\varphi_0$ [deg]} & \\
			\cline{8-11}
			Aperture & Array & TRMs & $d/\lambda$ & $\theta_0$ [deg] & Metric & beam & 0 & 45 & 90 & 135 & Time \\
			\hline
			\multirow{4}{*}{$50\times50$}
			& \multirow{2}{*}{\cite{yang2024tiling}} & \multirow{2}{*}{1249} & \multirow{2}{*}{0.50} & \multirow{2}{*}{45}
			& PSL & $-20$ & $-20$ & $-20$ & $-20$ & $-20$ & \multirow{2}{*}{28 h} \\
			& & & & & $D$ & 38.66 & 34.48 & 34.09 & 34.39 & 34.06 & \\
			\cline{2-12}
			& \multirow{2}{*}{Proposed} & \multirow{2}{*}{1250} & \multirow{2}{*}{0.50} & \multirow{2}{*}{45}
			& PSL & $-20$ & $-20$ & $-20$ & $-20$ & $-20$ & \multirow{2}{*}{840 s} \\
			& & & & & $D$ & 38.63 & 34.89 & 33.97 & 34.82 & 33.97 & \\
			\hline
			\multirow{4}{*}{$80\times80$}
			& \multirow{2}{*}{\cite{Anselmi2026}} & \multirow{2}{*}{3200} & \multirow{2}{*}{0.52} & \multirow{2}{*}{60}
			& PSL & $-13.30$ & $-12.52$ & -- & $-12.59$ & -- & \multirow{2}{*}{46 h} \\
			& & & & & $D$ & 43.37 & 35.69 & -- & 35.95 & -- & \\
			\cline{2-12}
			& \multirow{2}{*}{Proposed} & \multirow{2}{*}{3200} & \multirow{2}{*}{0.52} & \multirow{2}{*}{60}
			& PSL & $-13.53$ & $-13.72$ & $-14.17$ & $-13.67$ & $-14.27$ & \multirow{2}{*}{3350 s} \\
			& & & & & $D$ & 43.34 & 37.95 & 35.77 & 37.93 & 35.77 & \\
			\hline
			\multirow{4}{*}{$100\times100$}
			& \multirow{2}{*}{\cite{yang2024tiling}} & \multirow{2}{*}{4981} & \multirow{2}{*}{0.50} & \multirow{2}{*}{45}
			& PSL & $-13.27$ & $-13.03$ & $-13.08$ & $-13.04$ & $-13.09$ & \multirow{2}{*}{$\approx$100 h} \\
			& & & & & $D$ & 44.94 & 40.08 & 39.45 & 39.98 & 39.64 & \\
			\cline{2-12}
			& \multirow{2}{*}{Proposed} & \multirow{2}{*}{5000} & \multirow{2}{*}{0.50} & \multirow{2}{*}{45}
			& PSL & $-13.31$ & $-13.28$ & $-13.08$ & $-13.10$ & $-13.05$ & \multirow{2}{*}{7050 s} \\
			& & & & & $D$ & 44.95 & 41.23 & 40.29 & 41.14 & 40.30 & \\
			\hline\hline
		\end{tabular}
		\begin{tablenotes}\footnotesize
			\item PSL [dB] and directivity $D$ [dBi] are given for the normal beam and for the beam
			scanned to $(\theta_0,\varphi_0)$, with $\theta_0=45^\circ$ for the $50\times50$ and
			$100\times100$ apertures and $\theta_0=60^\circ$ for the $80\times80$ aperture (only
			$\varphi_0=0^\circ$ and $90^\circ$ for the reference). The proposed time is the total
			hybrid time (dataset, training, and search) plus one convex optimization.
		\end{tablenotes}
	\end{threeparttable}
\end{table*}

% Figures are placed right after the table, in the same float group,
% so they appear below it (figure* floats at the top of the page, in source order).
\begin{figure}[t]
	\centering
	\includegraphics[width=\columnwidth]{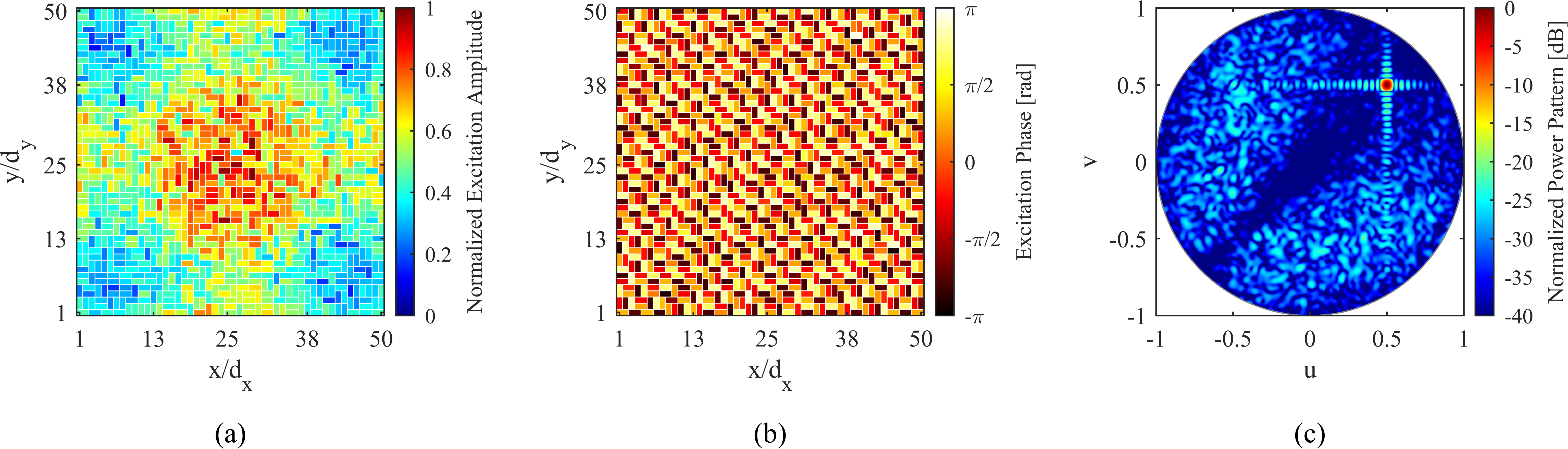}\\[1mm]
	\includegraphics[width=\columnwidth]{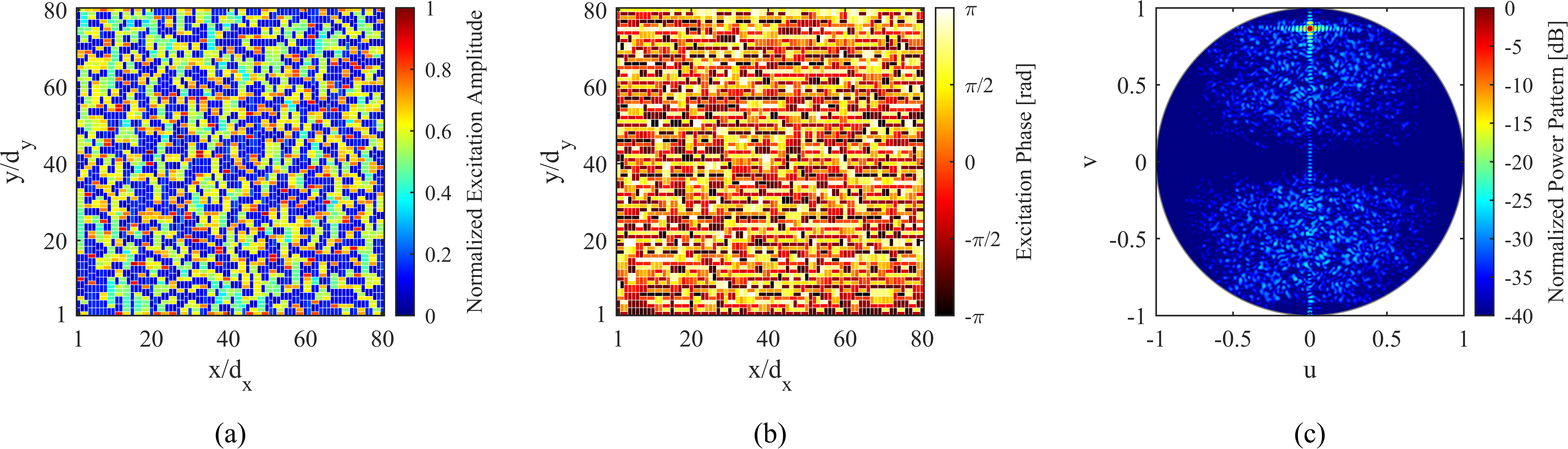}\\[1mm]
	\includegraphics[width=\columnwidth]{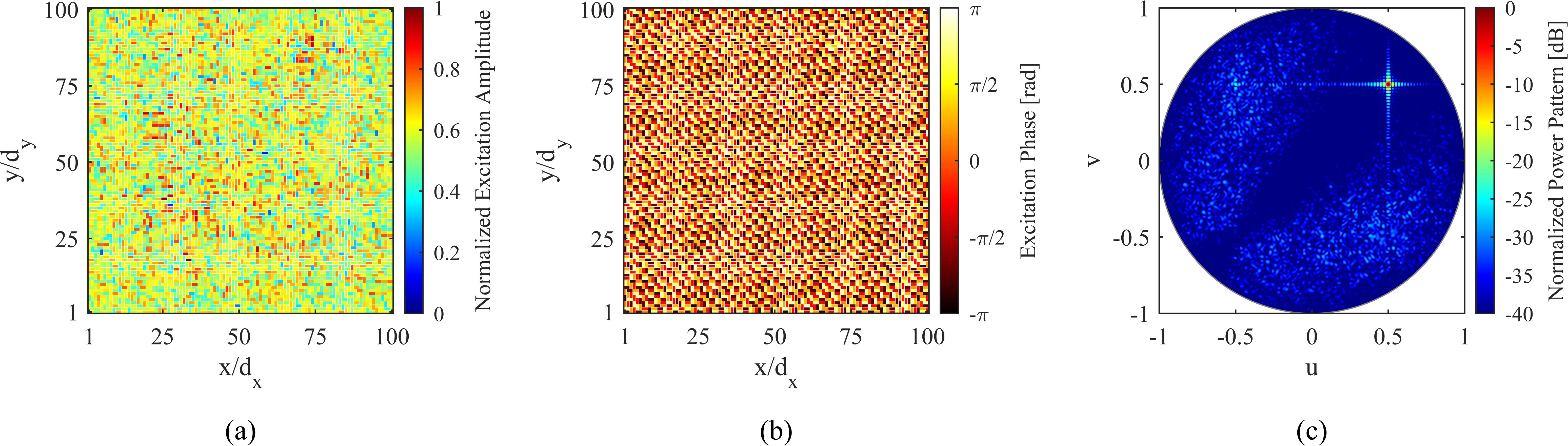}
	\caption{Convex amplitude--phase optimization of the $50\times50$ array scanned to $(45^\circ,45^\circ)$ (top), the $80\times80$ array scanned to $(60^\circ,90^\circ)$ (middle), and the $100\times100$ array scanned to $(45^\circ,45^\circ)$ (bottom). For each array: (a) normalized excitation amplitude, (b) excitation phase, and (c) normalized power pattern in the $(u,v)$ plane.}
	\label{fig:ap_excitation}
\end{figure}

\begin{figure}[t]
	\centering
	\includegraphics[width=0.32\columnwidth]{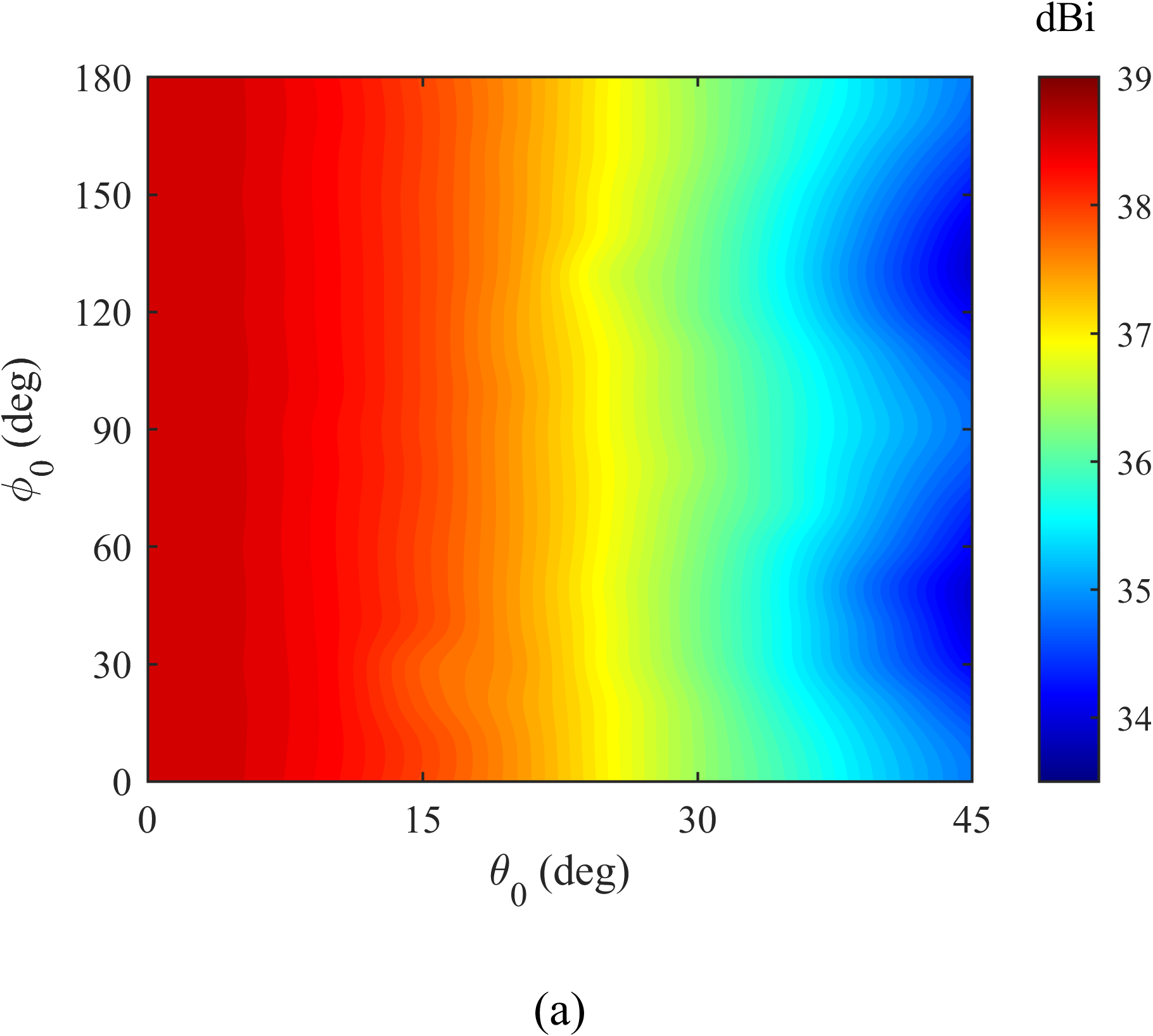}\hfill
	\includegraphics[width=0.32\columnwidth]{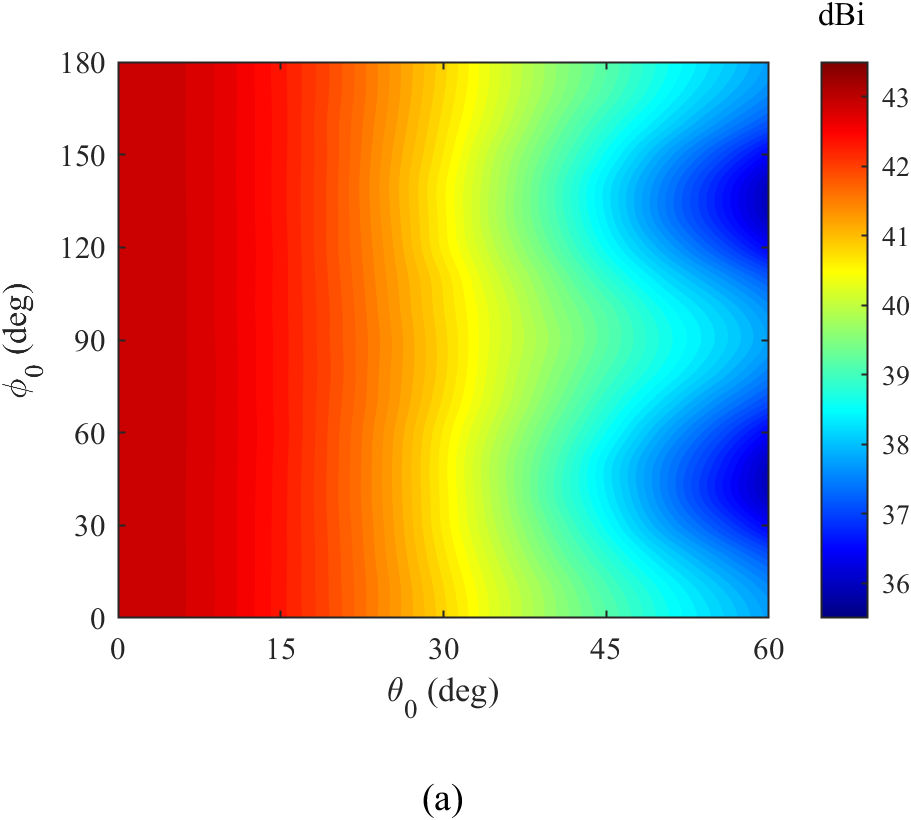}\hfill
	\includegraphics[width=0.32\columnwidth]{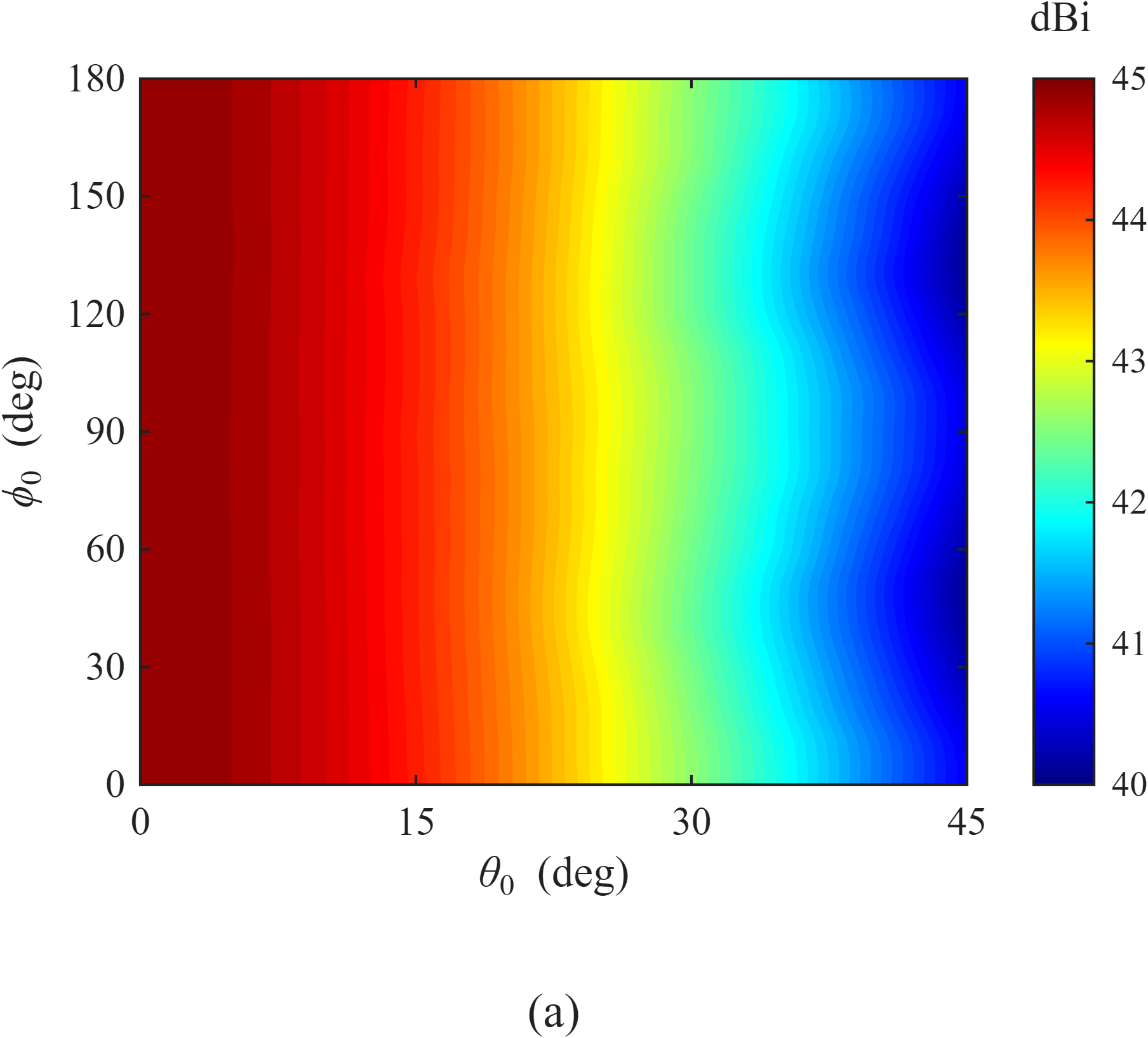}
	\caption{Directivity versus scan angle $(\theta_0,\varphi_0)$ for the $50\times50$ (left), $80\times80$ (middle), and $100\times100$ (right) arrays. The maps were computed at 172 scan directions and interpolated elsewhere. The scan range of the $80\times80$ array extends to $\theta_0=60^\circ$.}
	\label{fig:ap_directivity}
\end{figure}

Table~\ref{tab:summary} compares the proposed arrays with the references. In all three cases the
number of TRMs is essentially the same, and the proposed approach gives equal or higher
directivity at a comparable PSL with a computation time that is more than an order of magnitude
shorter.

\textit{$50\times50$:} The PSL is $-20$~dB for both designs. The proposed array has about
$0.4$~dB higher directivity at $\varphi_0=0^\circ$ and $90^\circ$ and about $0.1$~dB lower at
$45^\circ$ and $135^\circ$ (mean gain $0.16$~dB), so the directivity is comparable, while the
time drops from $28$~h to $840$~s.

\textit{$80\times80$:} This is the most demanding case, since the tiling was optimized for
$45^\circ$ and not re-optimized for the $60^\circ$ scan of \cite{Anselmi2026}. The normal-beam
directivity is the same (within $0.03$~dB), but the proposed array has $2.26$ and $1.98$~dB
higher directivity at $\varphi_0=0^\circ$ and $90^\circ$ and a PSL lower by $1.2$ and $1.08$~dB,
i.e., a scan loss of about $5.4$~dB versus $7.4$--$7.7$~dB. The excitation optimization
therefore compensates for a tiling not matched to the scan angle, so the tiling search need not
be repeated for every scan condition. The time drops from $46$~h to $3350$~s.

\textit{$100\times100$:} The normal-beam directivity is identical ($44.95$ versus $44.94$~dBi)
and the PSL differs by at most $0.25$~dB (slightly worse only at $\varphi_0=135^\circ$, $-13.05$
versus $-13.09$~dB). The directivity is higher by $1.15$, $0.84$, $1.16$, and $0.66$~dB at
$\varphi_0=0^\circ$, $45^\circ$, $90^\circ$, and $135^\circ$ ($0.95$~dB on average), while the
time drops from about $100$~h to $7050$~s.

The time saved is about $28$, $45$, and $98$~hours for the three cases. Since each of the five
beams in Table~\ref{tab:summary} needs its own convex optimization, the total for all five is
about $900$, $3800$, and $8000$~s, giving speed-ups of about $110$, $43$, and $45$, which is
still more than an order of magnitude. As the references and scan angles differ between cases,
these numbers show that the references become impractical as the aperture grows, not a general
trend with aperture size.

Fig.~\ref{fig:ap_excitation} shows the optimized excitations and patterns: each pattern has a
single main beam with the sidelobes held at the specified level over the visible hemisphere. The
$80\times80$ amplitude spreads over the whole range, consistent with the tiling not being matched
to $60^\circ$.

Fig.~\ref{fig:ap_directivity} shows the directivity over the scan range
($0^\circ\le\theta_0\le45^\circ$ for $50\times50$ and $100\times100$, $0^\circ\le\theta_0\le60^\circ$
for $80\times80$, and $0^\circ\le\varphi_0\le180^\circ$). From broadside to the largest scan
angle, the directivity loss is $3.7$--$4.7$~dB for $50\times50$, $5.4$--$7.6$~dB for
$80\times80$ at $60^\circ$, and $3.7$--$4.7$~dB for $100\times100$, of which $1.5$~dB
($45^\circ$) and $3$~dB ($60^\circ$) is the intrinsic $\cos\theta_0$ loss. The dependence on
$\varphi_0$ is negligible near broadside and reaches about $0.9$, $2.2$, and $0.9$~dB at the
largest angle. The $80\times80$ map stays smooth up to $60^\circ$, beyond the $45^\circ$ for which
the tiling was optimized, so the convex optimization extends the useful scan range of a tiling
beyond its design angle.
\subsection{Application to Dual-Band LEO Satellite Terminal Arrays}
\label{subsec:leo_application}

LEO broadband constellations depend on electronically steered user terminals
that need a wide scan range, good circular polarization over that range, and a thin, inexpensive
aperture. In the Ku-band, non-geostationary systems receive in 10.7--12.75~GHz and transmit in
14.0--14.5~GHz, with linear or circular polarization allowed \cite{ETSI_EN303980,ETSI_TR103399}.
Circularly polarized phased arrays with wide-angle scanning \cite{SIW_Kband} and shared-aperture
arrays covering two satellite bands \cite{Li2024CKu} have been reported. A single aperture for
both bands removes a second antenna from the terminal, but it is harder to design.

Here, the framework is applied to a $36\times36$ array that receives at 11.2~GHz and transmits at
14.0~GHz with circular polarization and scans up to $\theta_{\max}=50^\circ$, a value set by the
application. The design frequencies lie inside the receive allocation and at the lower edge of
the transmit allocation. The requirements are a thin dual-band circularly polarized element, a
common element spacing suitable for both bands, and a peak SLL of $-13.5$~dB or lower over the
scan range with as few control points as possible.

Domino tiling suits this application for three reasons. First, with one control point per tile,
the 1296 elements need only 648 control points per band. Second, the irregular arrangement of the
tiles breaks the periodicity of the tile phase centers, so quantization lobes do not add up at
the same angles as in a regular subarray layout; irregular tilings of large apertures have been
studied for LEO communication, with feeding-network constraints, and with simplified 3D-printed
beamforming networks \cite{Anselmi2026}. Third, all elements keep the same
orientation and only the grouping of the excitations is irregular, so the polarization purity is
set by the axial ratio of the element, unlike the sequential-rotation and phase-compensation
approaches of \cite{Lin2023,Yang2024}, which rotate the tiles.

Irregular arrays suffer from gain drop and high sidelobes at large scan angles; unequal phase
feeding has been used to extend the scan range of an irregular array from $\pm35^\circ$ to
$\pm65^\circ$ \cite{Ma2024}. Here, the convex amplitude--phase optimization of the previous
subsection plays the same role on top of the fixed tiling.

The two bands are a factor of 1.25 apart, so one element must support two close but independent
resonances, and the shared element spacing must satisfy two electrical conditions: the upper
band limits it through grating lobes, while in the lower band the elements are electrically
closer, which increases mutual coupling. We therefore first characterize the element in both
bands, including its axial ratio over the whole hemisphere, and then study the array.

\subsubsection{Dual-Band Circularly Polarized Radiating Element}
\label{subsubsec:dualband_element}

The element is a center-fed dual-band circularly polarized stacked-patch antenna. Two patches provide the two close resonances at 11.2 and 14.0~GHz: a probe-fed radiation patch and a parasitic patch above it. Following the compact dual-band GPS antenna of Sun \emph{et al.}\ \cite{Sun2023}, both patches are redesigned for the Ku-band with an annular gap and a pair of stubs that generate circular polarization with a single feed.

To improve polarization purity with a compact footprint, we add a shorted annular ring inspired by the reduced-surface-wave (RSW) concept, which has been reported to give a higher RHCP-to-LHCP ratio and lower backside radiation than choke-ring antennas \cite{Basilio2007}. Unlike the classical shorted annular patch, our ring is a circular row of plated shorting holes that surrounds the stacked patches, and its radius was tuned together with the patch dimensions in full-wave simulation.

\begin{figure}[t]
	\centering
	\begin{subfigure}[b]{0.8\columnwidth}
		\centering
		\includegraphics[width=\linewidth]{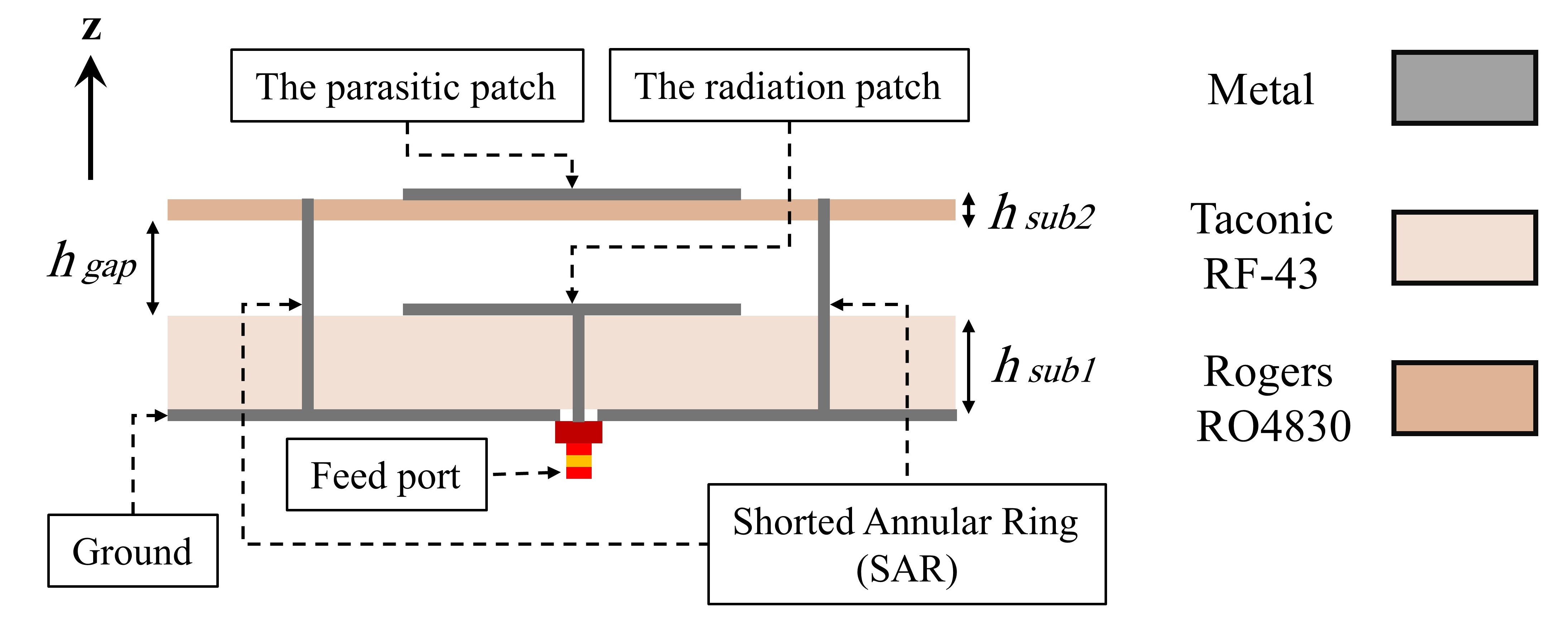}
		\caption{}
		\label{fig:ant_side}
	\end{subfigure}
	
	\vspace{3mm}
	
	\begin{subfigure}[b]{0.42\columnwidth}
		\centering
		\includegraphics[width=\linewidth]{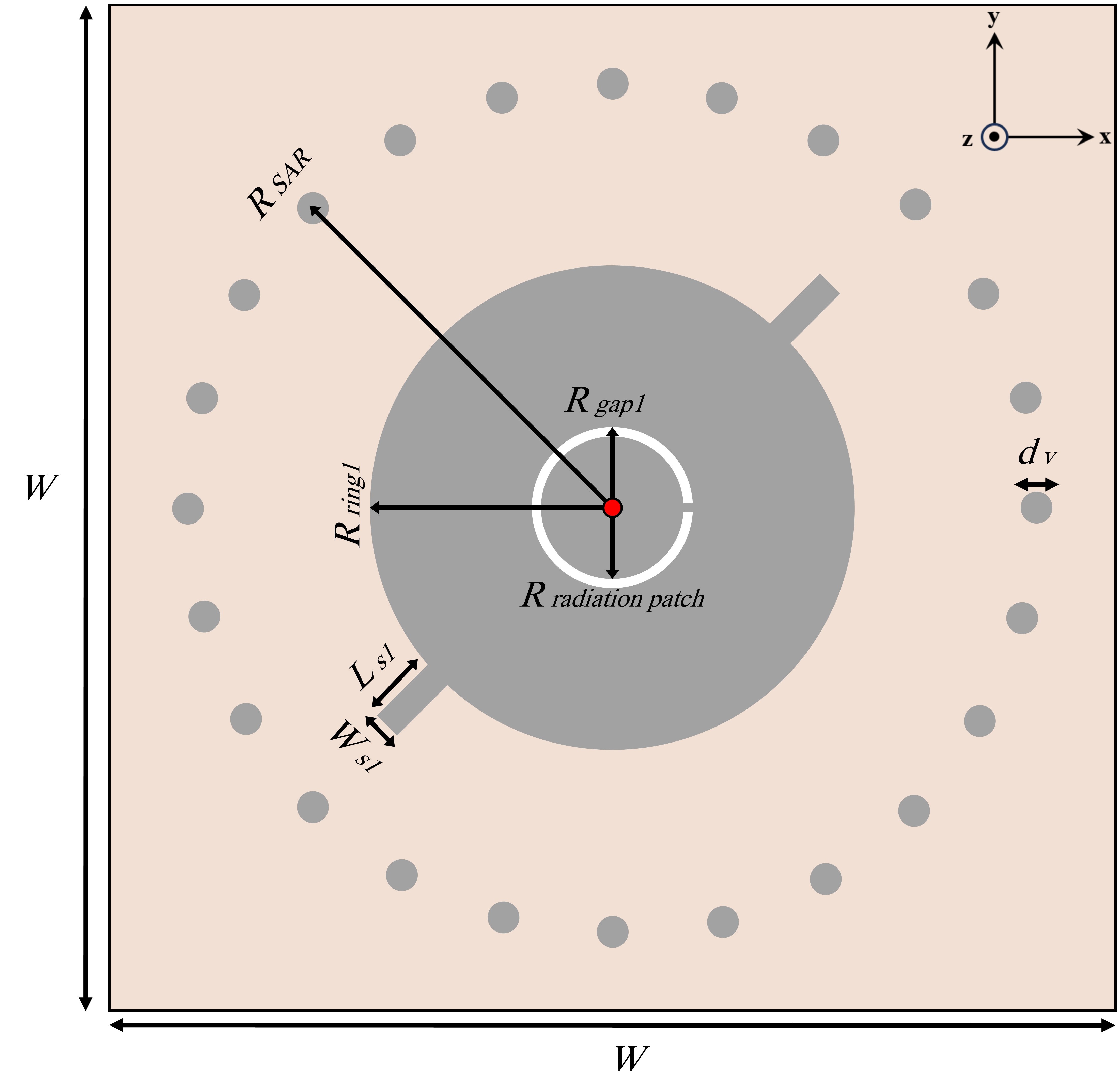}
		\caption{}
		\label{fig:ant_rad}
	\end{subfigure}
	\begin{subfigure}[b]{0.42\columnwidth}
		\centering
		\includegraphics[width=\linewidth]{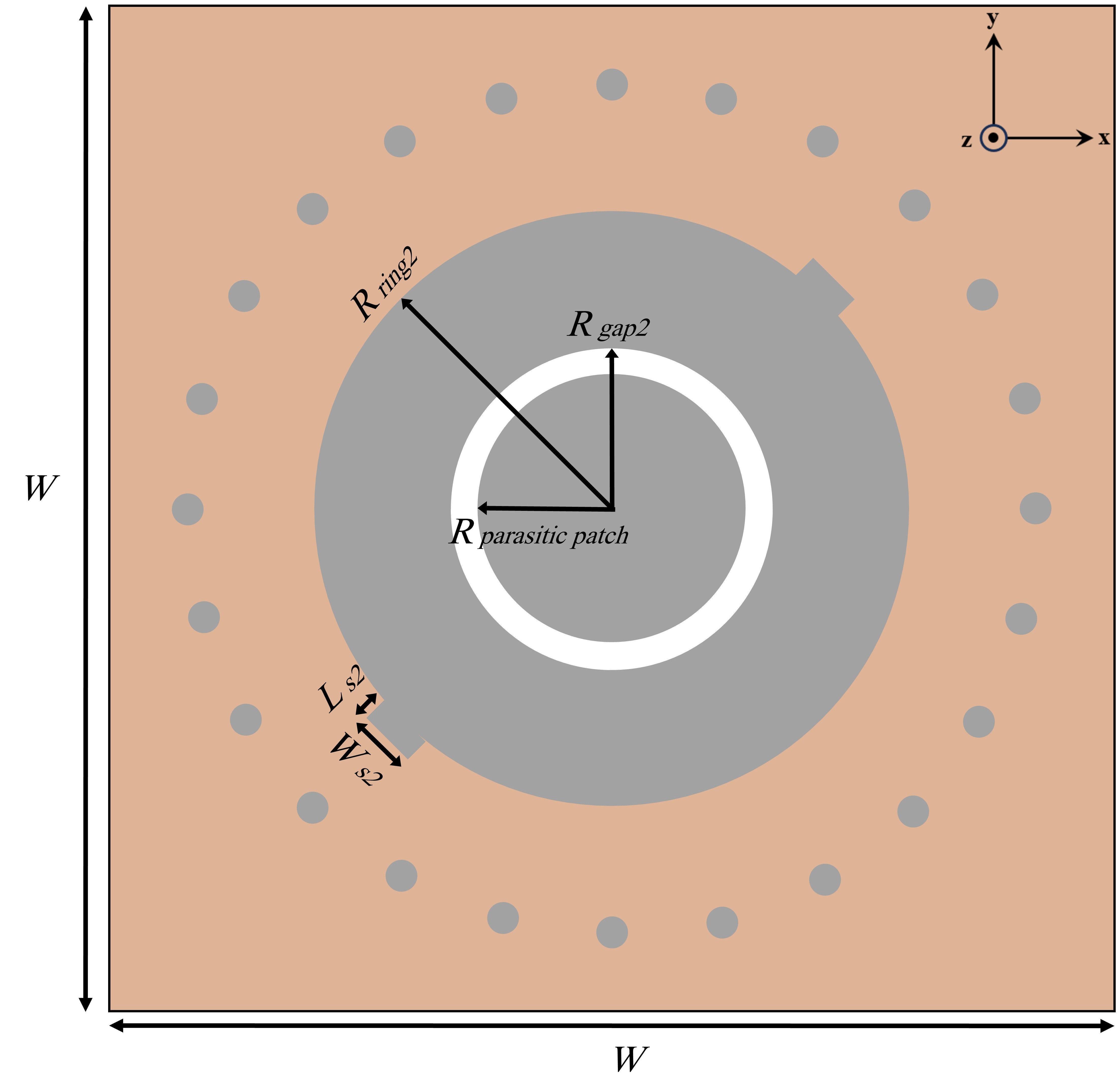}
		\caption{}
		\label{fig:ant_par}
	\end{subfigure}
	
	\caption{Geometry of the proposed dual-band antenna: (a) side view, (b) top view of the radiation patch layer, and (c) top view of the parasitic patch layer.}
	\label{fig:antenna_structure}
\end{figure}

The geometry is shown in Fig.~\ref{fig:antenna_structure}. A radiation patch is printed on a
Taconic RF-43 substrate of thickness $h_{\mathrm{sub1}}$ and fed at its center by a probe. A
parasitic patch is printed on a thin Rogers RO4830 layer of thickness $h_{\mathrm{sub2}}$ above
it, separated by a gap $h_{\mathrm{gap}}$. The total stack thickness is
$h_{\mathrm{sub1}}+h_{\mathrm{sub2}}+h_{\mathrm{gap}}=1.003$~mm, i.e., $0.037\lambda_0$ at
11.2~GHz. Each patch has a central disc and an outer ring separated by a narrow annular gap, and
the outer ring carries two opposite rectangular stubs that perturb the orthogonal modes and
generate circular polarization with a single feed. The shorted ring uses $N_v=36$ holes of
diameter $d_v=0.2$~mm at radius $R_{\mathrm{SAR}}=5.15$~mm, so its outer diameter
($10.5$~mm) is smaller than the element spacing and neighboring rings do not overlap. The final
dimensions are listed in Table~\ref{tab:params}.

\begin{table}[t]
	\centering
	\caption{Parameters of the Proposed Dual-Band Antenna}
	\label{tab:params}
	\footnotesize
	\setlength{\tabcolsep}{2pt}
	\renewcommand{\arraystretch}{1.15}
	\begin{tabular}{cccccccc}
		\hline\hline
		Par. & \shortstack{Values\\(mm)} & Par. & \shortstack{Values\\(mm)} & Par. & \shortstack{Values\\(mm)} & Par. & \shortstack{Values\\(mm)} \\
		\hline
		$W$        & 11.500 & $R_{SAR}$                    & 5.150 & $R_{ring1}$ & 3.205 & $R_{gap2}$  & 2.151 \\
		$h_{sub1}$ & 0.355  & $d_v$                        & 0.200 & $L_{s1}$    & 0.725 & $R_{ring2}$ & 3.937 \\
		$h_{sub2}$ & 0.127  & $R_{\text{radiation patch}}$ & 0.943 & $W_{s1}$    & 0.366 & $L_{s2}$    & 0.333 \\
		$h_{gap}$  & 0.521  & $R_{gap1}$                   & 1.065 & $R_{\text{parasitic patch}}$ & 1.774 & $W_{s2}$ & 0.776 \\
		\hline\hline
	\end{tabular}
\end{table}

\begin{figure}[!t]
	\centering
	\begin{subfigure}[b]{0.6\columnwidth}
		\centering
		\includegraphics[width=\linewidth]{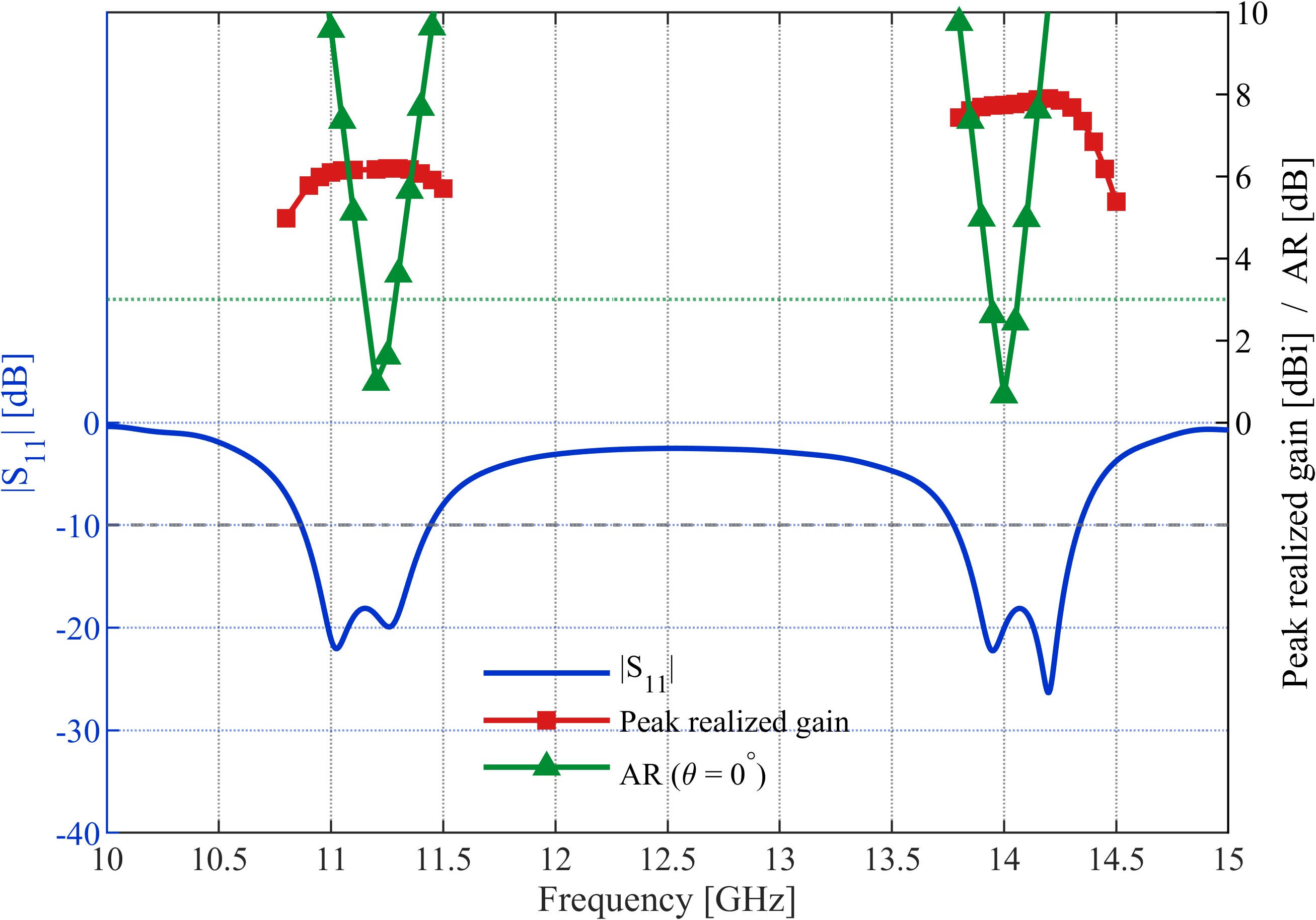}
		\caption{}
		\label{fig:A_s11}
	\end{subfigure}
	
	\vspace{2mm}
	
	\begin{subfigure}[b]{0.45\columnwidth}
		\centering
		\includegraphics[width=\linewidth]{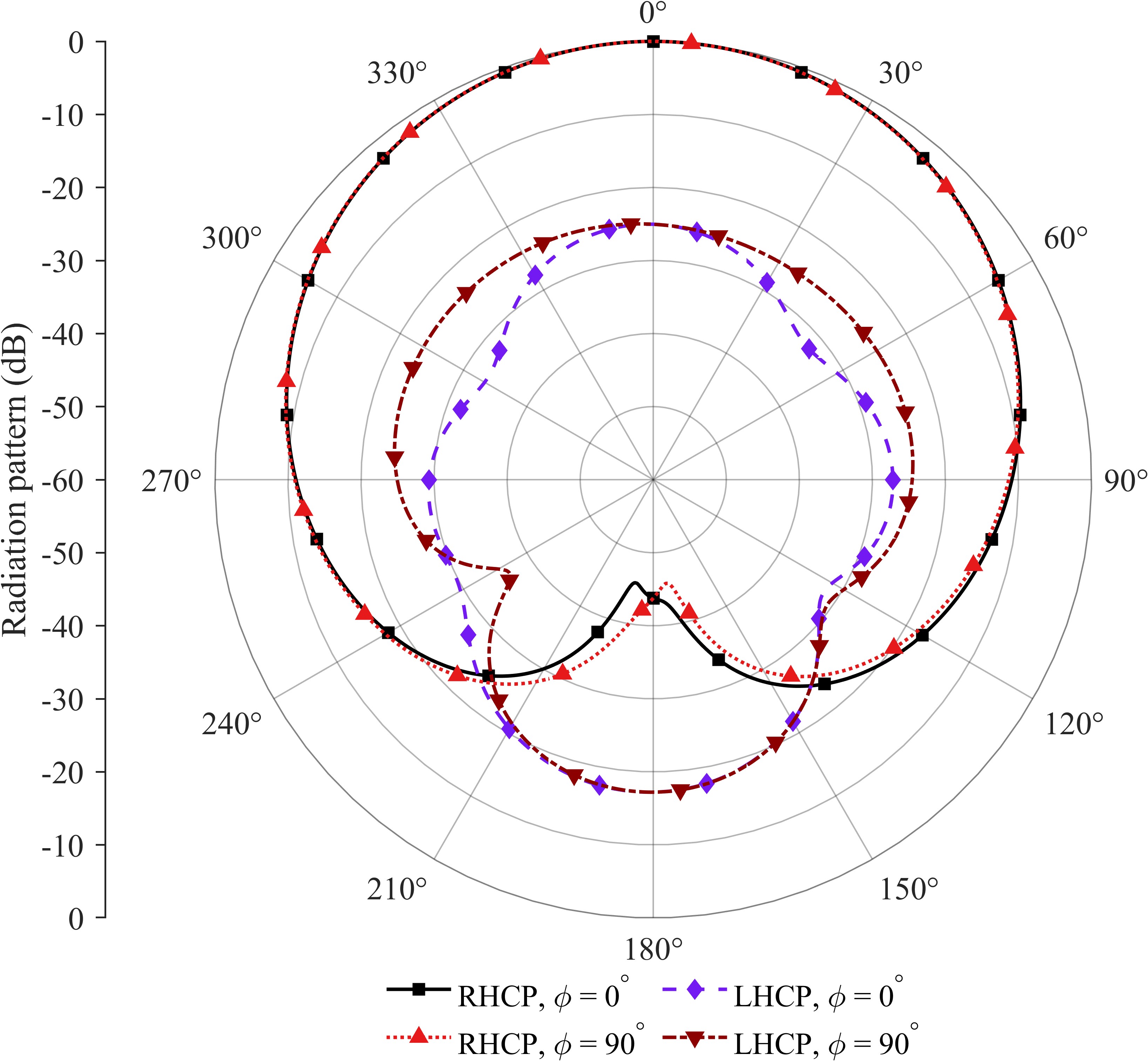}
		\caption{}
		\label{fig:B_polar11}
	\end{subfigure}\hfill
	\begin{subfigure}[b]{0.45\columnwidth}
		\centering
		\includegraphics[width=\linewidth]{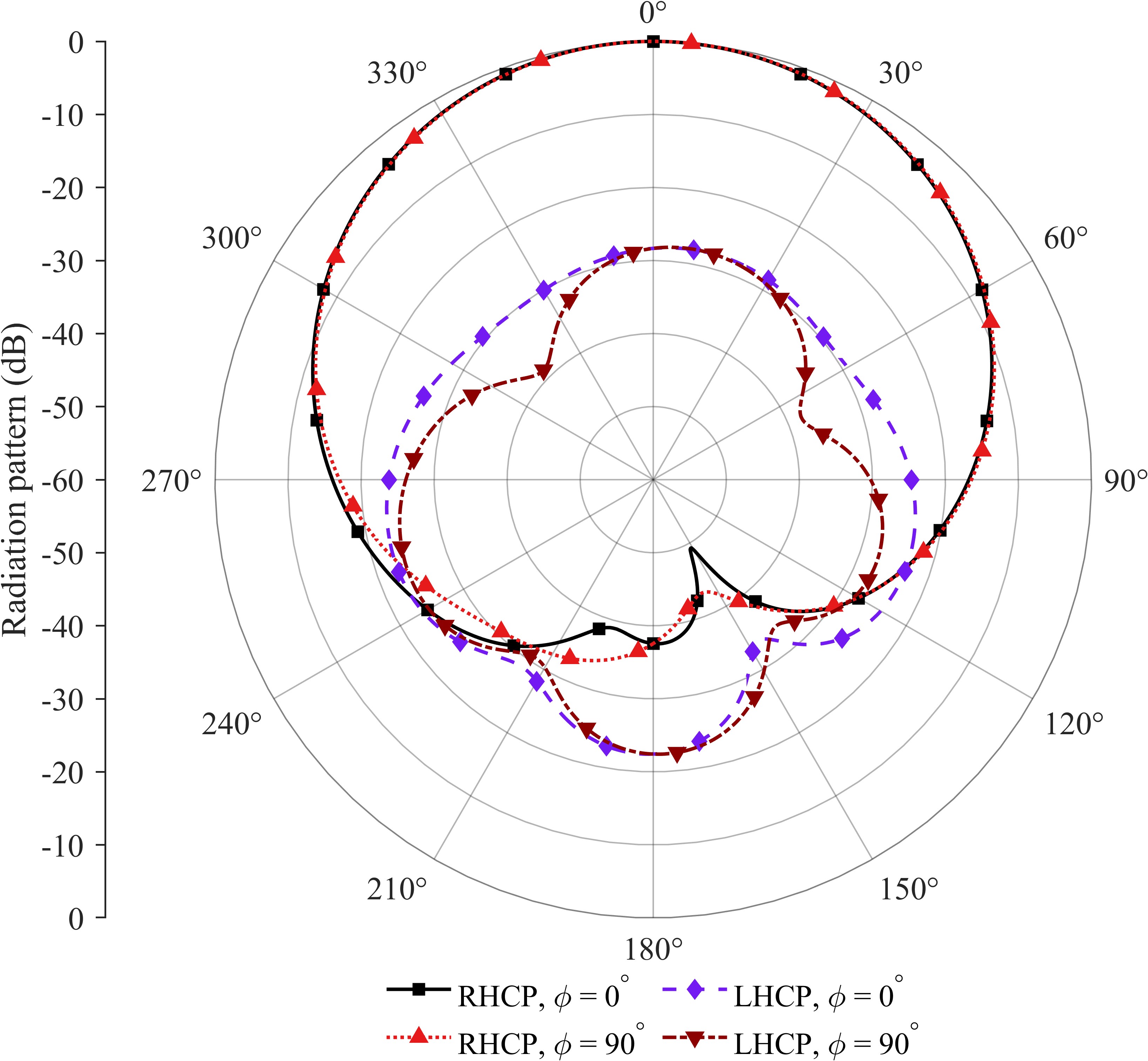}
		\caption{}
		\label{fig:C_polar14}
	\end{subfigure}
	
	\caption{Simulated results: (a) $|S_{11}|$, peak realized gain, and AR at $\theta=0^\circ$ versus frequency, and RHCP/LHCP radiation patterns at (b) 11.2\,GHz and (c) 14.0\,GHz.}
	\label{fig:results}
\end{figure}

The simulated results are shown in Fig.~\ref{fig:results}. The reflection coefficient shows two resonances at 11.2 and 14.0~GHz, below $-15$~dB at both. At broadside, the peak realized gain is 5.86 and 7.48~dBic, the axial ratio stays below 3~dB over 150 and 100~MHz, and LHCP is suppressed by 29.5 and 27.5~dB in the lower and upper bands, with half-power beamwidths of about $92^\circ$ and $76^\circ$. The design is therefore dual-band with narrow axial-ratio bandwidths.

\begin{figure}[!t]
	\centering
	\begin{subfigure}[b]{0.45\columnwidth}
		\centering
		\includegraphics[width=\linewidth]{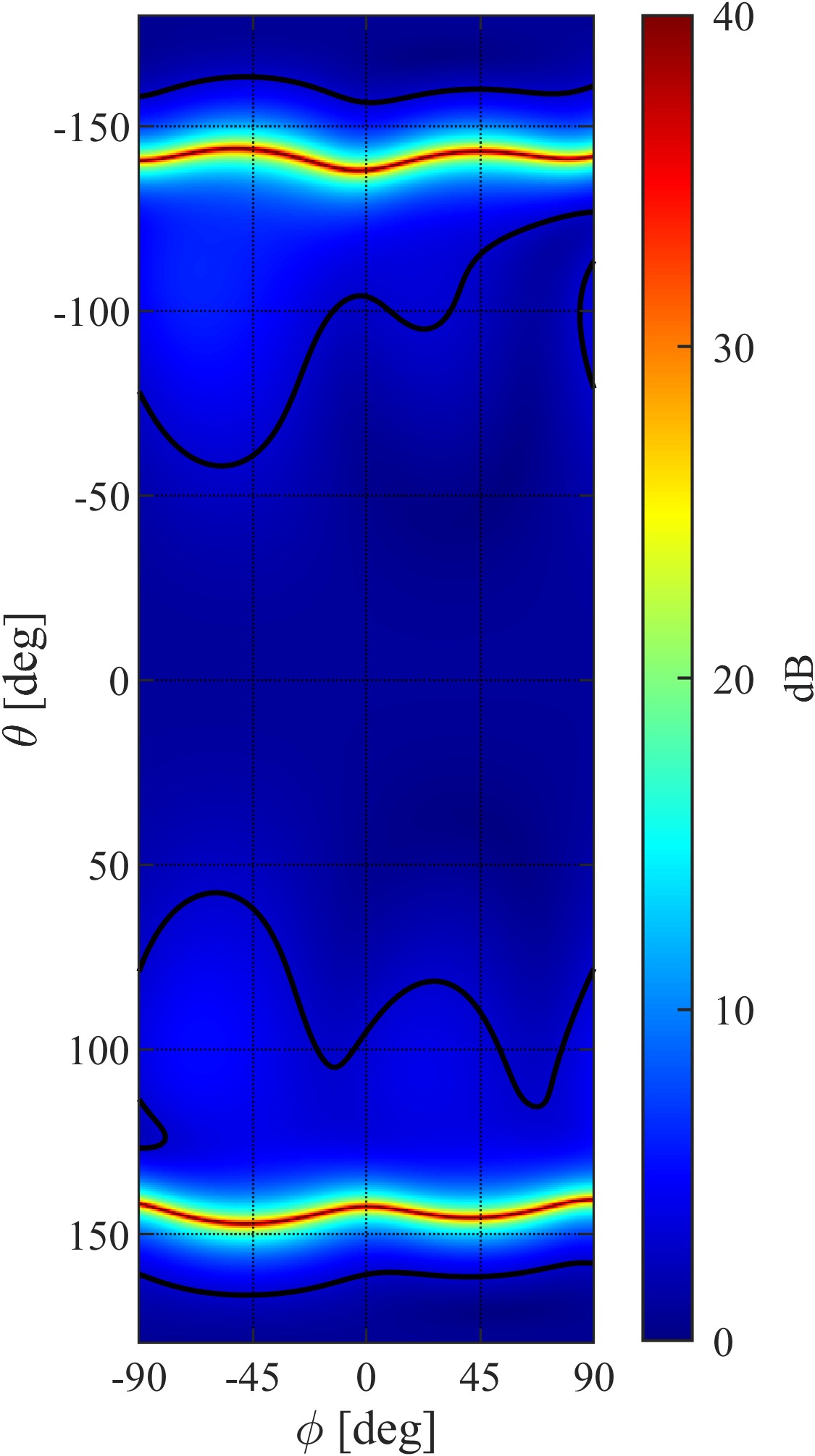}
		\caption{}
		\label{fig:AR_11200}
	\end{subfigure}\hfill
	\begin{subfigure}[b]{0.45\columnwidth}
		\centering
		\includegraphics[width=\linewidth]{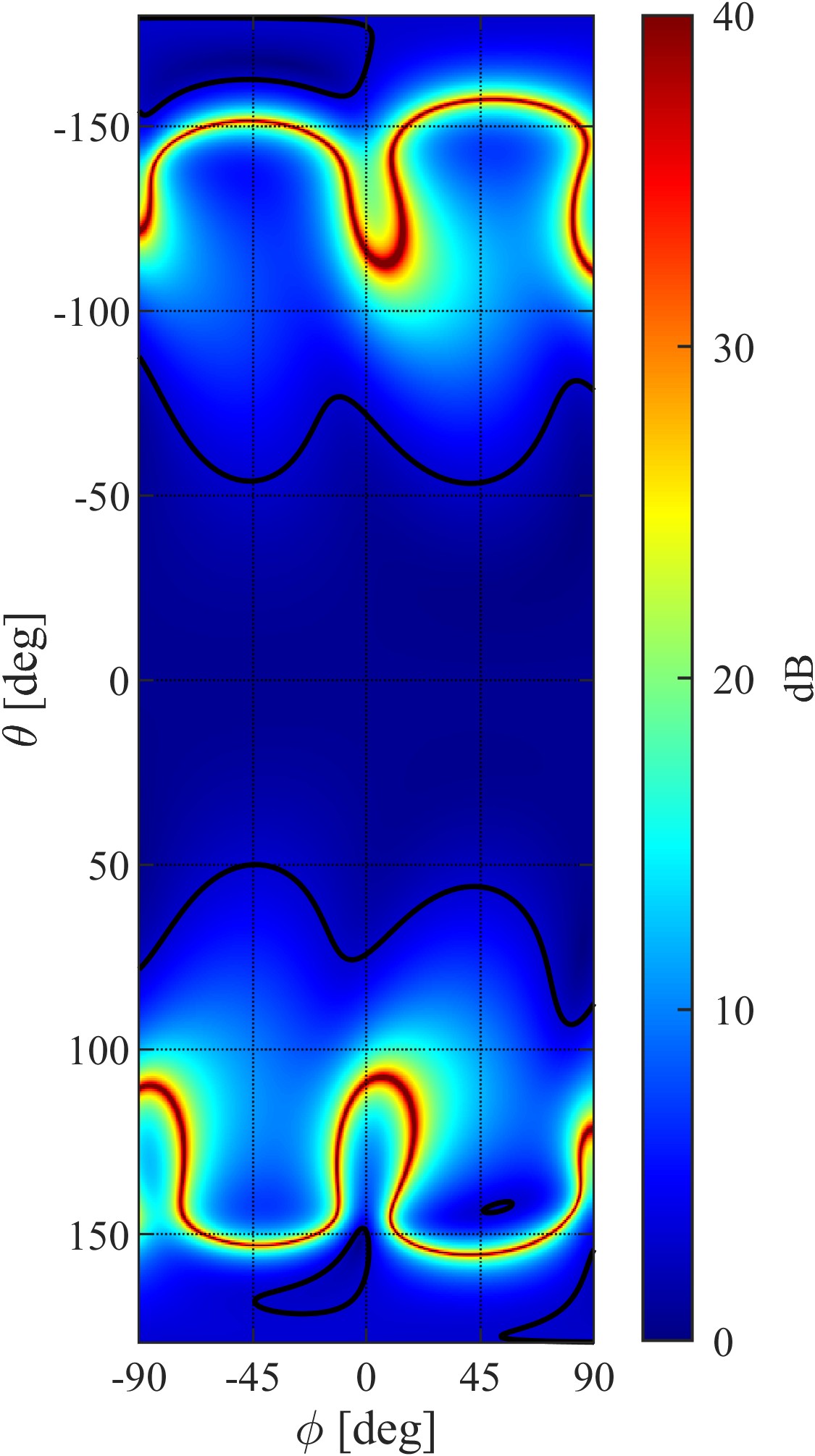}
		\caption{}
		\label{fig:AR_14000}
	\end{subfigure}
	\caption{Simulated axial ratio of the proposed antenna at (a) 11.2~GHz and (b) 14.0~GHz. The black contour indicates AR\,=\,3\,dB.}
	\label{fig:AR_maps}
\end{figure}

Fig.~\ref{fig:AR_maps} shows the axial-ratio distributions at the two frequencies, with the black
contours marking AR\,=\,3\,dB. In both bands the axial ratio stays below 3~dB up to
$\theta=50^\circ$. The lower band has a wider region satisfying the criterion, whereas the upper
band varies more with azimuth. Since all elements have the same orientation, this angular
dependence directly determines the polarization performance of the scanned array.

\subsubsection{Array Configuration and Element Spacing}
\label{subsubsec:dualband_array}

We consider a $36\times36$ planar array of 648 dominoes, each formed by two adjacent elements and
driven by one complex weight per band; the same tiling is used in both bands. Since the
optimization acts on the array factor, the simulated element far field is applied afterwards to
obtain the realized gain and the RHCP and LHCP patterns. Scanning is studied in the
$\phi_0=0^\circ$ plane for $\theta_0=0^\circ$--$50^\circ$ in $10^\circ$ steps.

The design is compared with the SCTIE-shaped irregular array of~\cite{liu2023sctie}, the closest
related work. That study considers a $36\times36$ array over $[0.9f_0,\,1.1f_0]$ (a ratio of
1.22, close to our $14.0/11.2=1.25$) with $d=0.5\lambda_h$, where $\lambda_h$ is the wavelength
at the upper edge. This spacing equals $0.5\lambda$ only at the upper edge and falls to about
$0.409\lambda$ at the lower edge, so we treated $d$ as a design parameter. Moreover,
in~\cite{liu2023sctie} the elements of one subarray are interleaved with those of others (every
other lattice site), so their feed lines are long, cross those of neighboring subarrays, and call
for a multilayer implementation. In our design the two elements of each domino are adjacent and
can be fed by a short local power divider.

Both bands share one aperture, so a common $d$ is needed. For a square lattice scanned in a
principal plane up to $\theta_{\max}$, grating lobes stay out of visible space when
\begin{equation}
	d \le \frac{\lambda}{1+\sin\theta_{\max}},
\end{equation}
which gives $0.566\lambda$ for $\theta_{\max}=50^\circ$, i.e., 15.2~mm at 11.2~GHz
($\lambda=26.77$~mm) and 12.1~mm at 14.0~GHz ($\lambda=21.41$~mm). The spacing was swept over
six values from 10.91 to 11.91~mm in steps of 0.2~mm, i.e., $0.407\lambda$--$0.445\lambda$ at
11.2~GHz and $0.510\lambda$--$0.556\lambda$ at 14.0~GHz. A large $d$ brings the upper band close
to the grating-lobe limit, while a small $d$ brings the lower band close to $0.4\lambda$. All
spacings exceed the outer diameter of the shorted ring, $2R_{\mathrm{SAR}}+d_v=10.5$~mm, so
neighboring rings never overlap; the substrate width $W$ of the simulated element does not
constrain the lattice.

The optimization was repeated for each spacing, and Fig.~\ref{fig:compare_d} shows the resulting
directivity and peak SLL. For a direct comparison with~\cite{liu2023sctie}, which assumes an
element pattern $e(\theta,\varphi)=\cos\theta$, the same pattern is used here instead of the
simulated one. In both bands the directivity increases with $d$, by about 0.75~dB at broadside
from $d=10.91$ to $11.91$~mm, in agreement with the growth of the aperture area
($20\log_{10}(11.91/10.91)=0.76$~dB). This advantage shrinks with scanning: at $\theta_0=50^\circ$
the spread is only 0.2--0.3~dB, and the largest spacing is no longer the best. The peak SLL
varies by only 0.1--0.2~dB with $d$, so the choice is governed by directivity and grating-lobe
margin. We selected $d=11.71$~mm ($0.437\lambda$ and $0.547\lambda$ at 11.2 and 14.0~GHz): its
directivity is within about 0.15~dB of that of $d=11.91$~mm at broadside and within 0.2~dB of the
best spacing at $50^\circ$, it leaves a larger grating-lobe margin in the upper band than
$d=11.91$~mm, and it keeps the lower band away from $0.4\lambda$. With this spacing, the
directivity (with the $\cos\theta$ pattern) decreases by 2.8~dB (lower band) and 4.6~dB (upper
band) from $\theta_0=0^\circ$ to $50^\circ$, so the upper-band advantage of 1.9~dB at broadside
vanishes at $50^\circ$.

Compared with the reference array of~\cite{liu2023sctie} (dotted lines), the proposed array has
a higher directivity at all scan angles in the lower band, by 0.3 to 1~dB. In the upper band it
is about 1~dB higher up to $\theta_0=30^\circ$, comparable at $40^\circ$, and about 1~dB lower at
$50^\circ$. The peak SLL is about 1~dB (lower band) and 0.25~dB (upper band) higher at
broadside, but 0.5 to 1~dB lower at large scan angles, i.e., from $\theta_0=40^\circ$ in the
lower band and from $20^\circ$ in the upper band. The design therefore trades a slightly higher
SLL near broadside for a lower SLL at large scan angles, gives a higher directivity except at
$50^\circ$ in the upper band, and uses a spacing more practical for the lower band and a
subarray structure that is easier to feed. Note that the arrays are evaluated with the array
factor and an element pattern, so mutual coupling is not included; a full-wave embedded-element
or finite-array simulation is needed to confirm the coupling-based arguments for the choice of
$d$.

\newcommand{\dlegitem}[3]{%
	\tikz[baseline=-0.6ex]{\draw[#1,line width=0.9pt] (0,0)--(0.7,0);
		\node[#1,fill=#1,#2,inner sep=0pt,minimum size=3pt] at (0.5,0) {};}%
	\,$d=#3$~mm}

\newcommand{\dlegend}{%
	\begin{tabular}{@{}l@{\hspace{1.2em}}l@{\hspace{1.2em}}l@{}}
		\dlegitem{c0}{circle}{10.91}  & \dlegitem{c1}{rectangle}{11.11} & \dlegitem{c2}{regular polygon,regular polygon sides=3}{11.31} \\[0.6ex]
		\dlegitem{c3}{diamond}{11.51} & \dlegitem{c4}{regular polygon,regular polygon sides=3,shape border rotate=180}{11.71} & \dlegitem{c5}{regular polygon,regular polygon sides=3,shape border rotate=270}{11.91}
\end{tabular}}

\begin{figure}[!t]
	\centering
	{\footnotesize\dlegend}\\[1ex]
	\subfloat[]{\includegraphics[width=0.45\columnwidth]{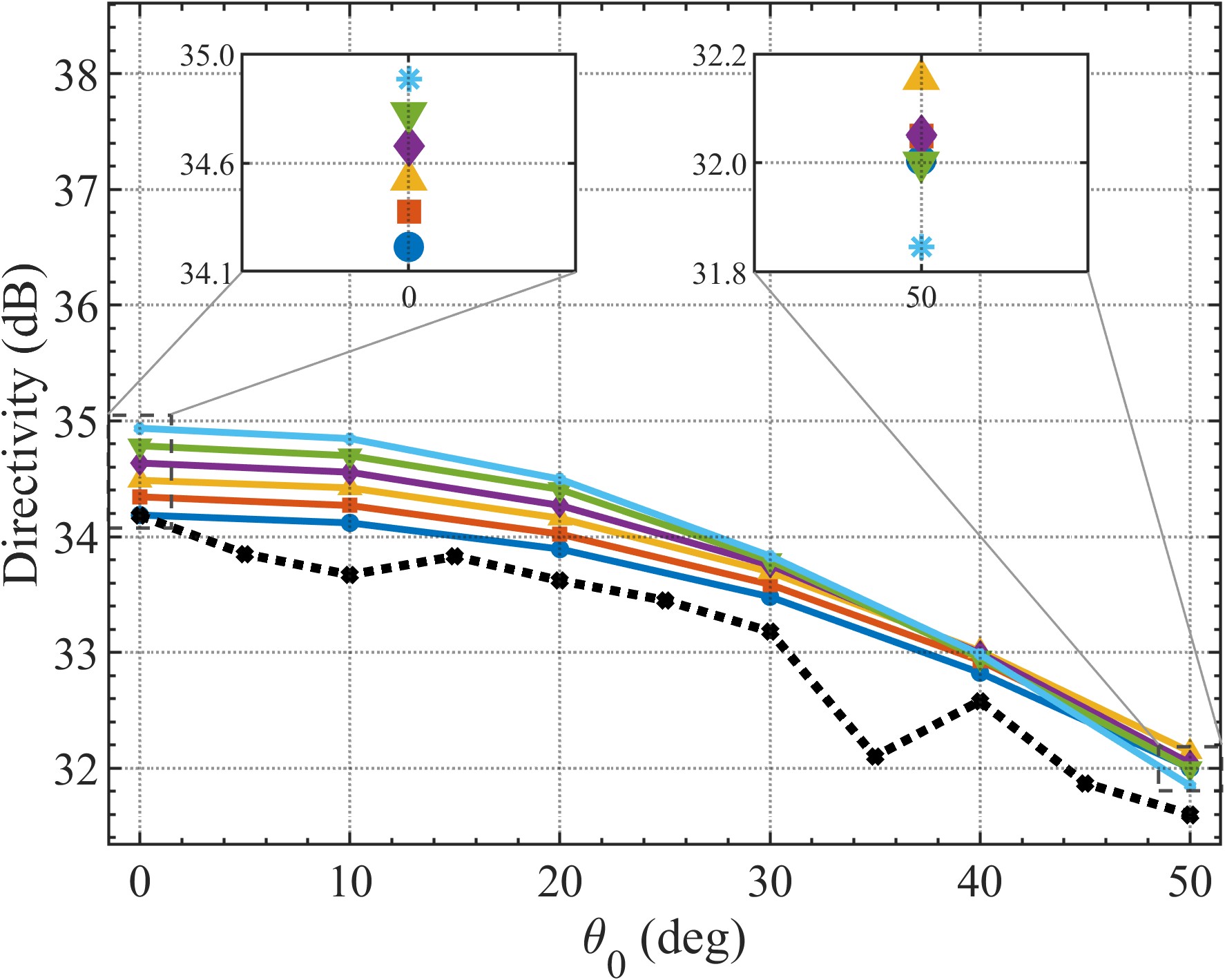}%
		\label{fig:cmp_DL}}
	\hfil
	\subfloat[]{\includegraphics[width=0.45\columnwidth]{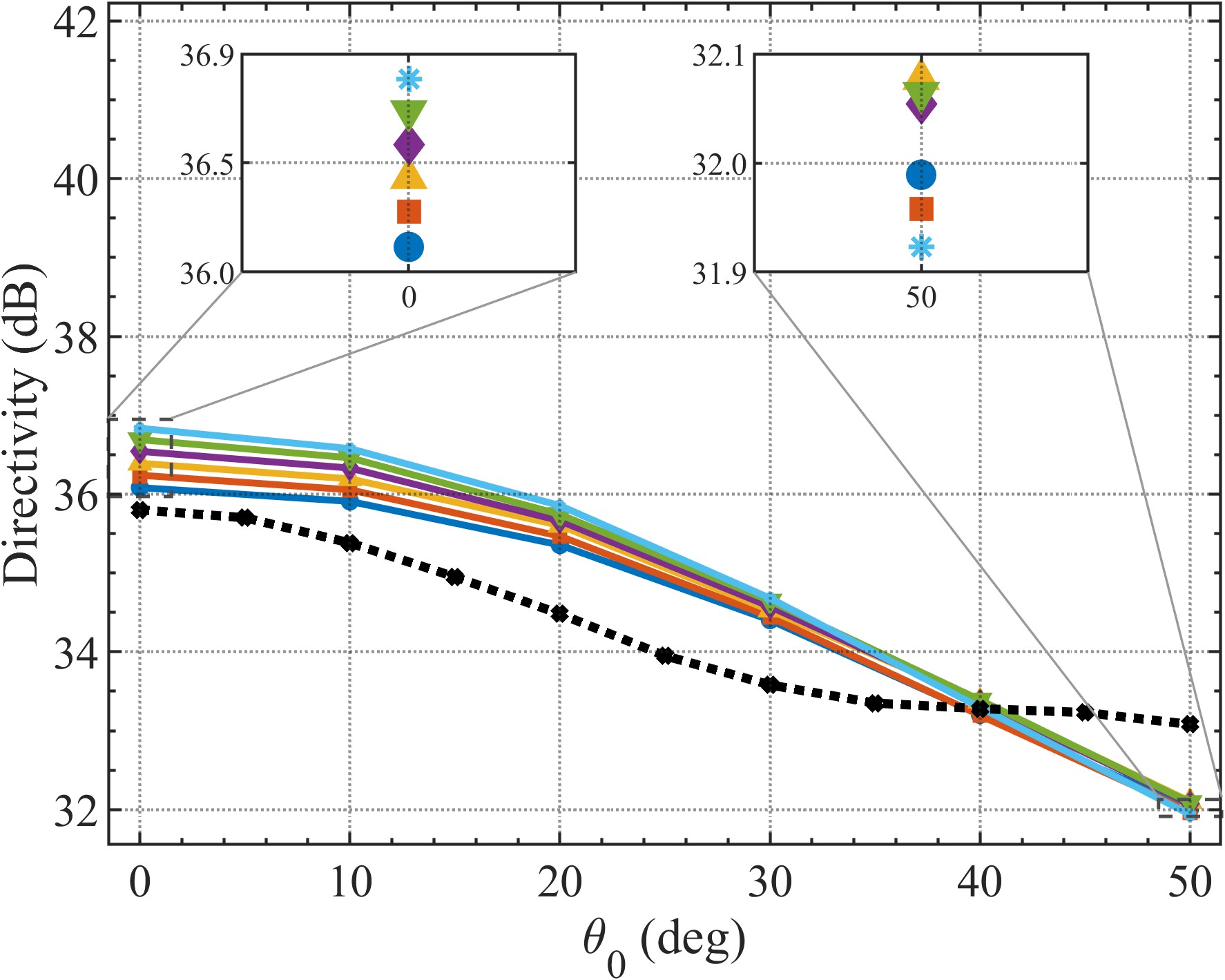}%
		\label{fig:cmp_DH}}
	\\
	\subfloat[]{\includegraphics[width=0.45\columnwidth]{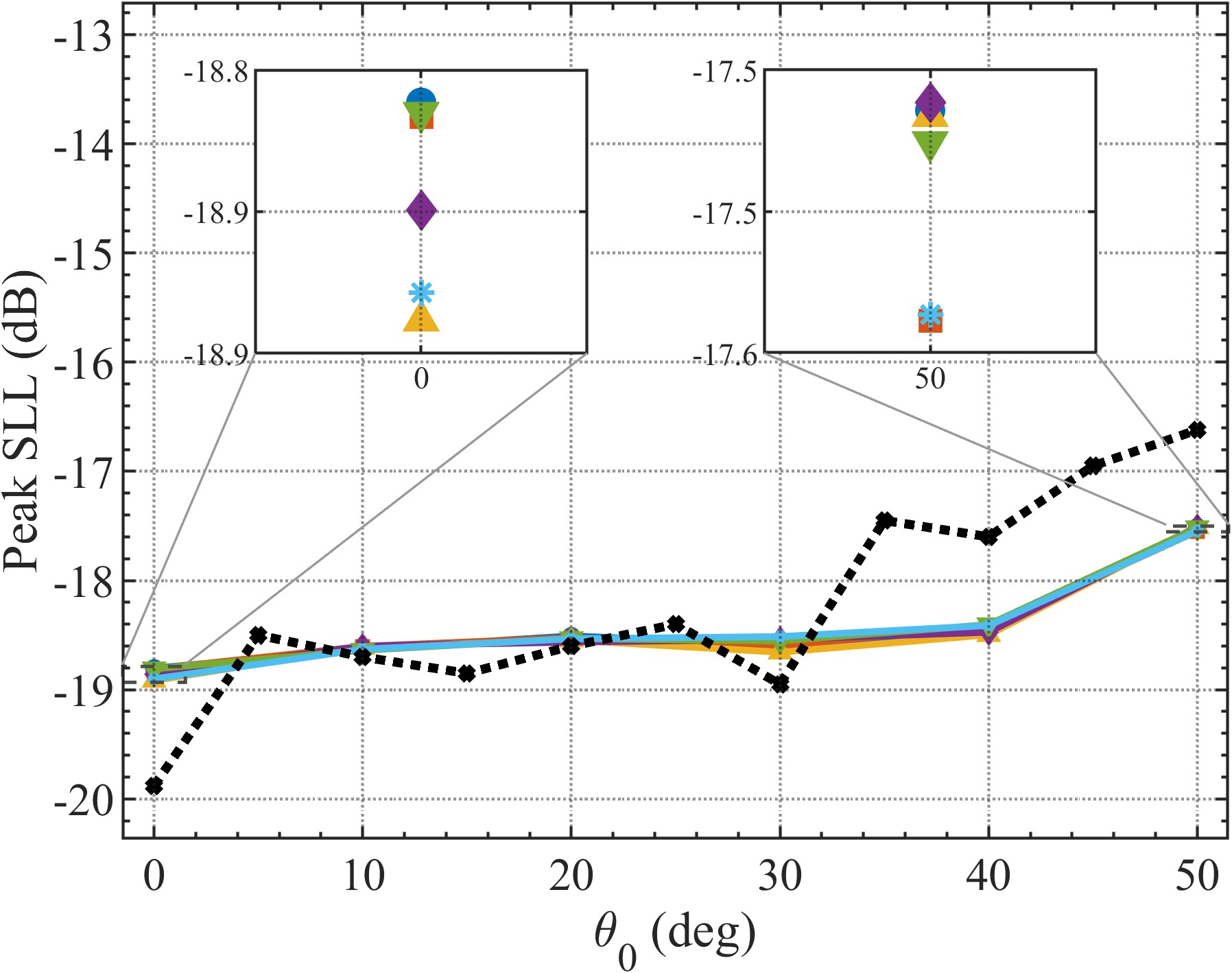}%
		\label{fig:cmp_SL}}
	\hfil
	\subfloat[]{\includegraphics[width=0.45\columnwidth]{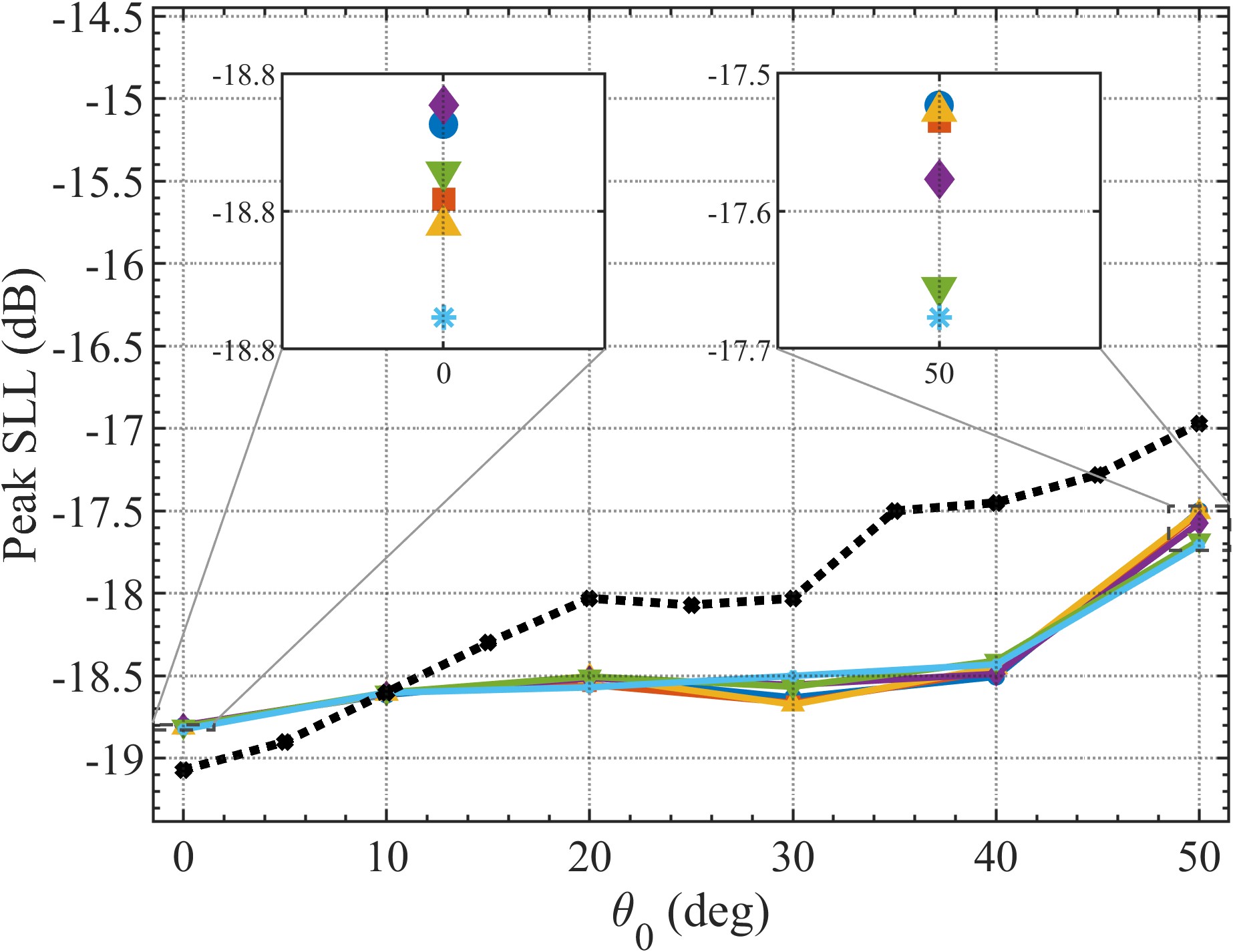}%
		\label{fig:cmp_SH}}
	\caption{Directivity and peak SLL versus $\theta_0$ in the $\phi_0=0^\circ$ plane for different element spacings $d$, assuming a $\cos\theta$ element pattern. (a)~Directivity, 11.2~GHz. (b)~Directivity, 14.0~GHz. (c)~Peak SLL, 11.2~GHz. (d)~Peak SLL, 14.0~GHz. The dotted lines show the corresponding results of the reference array in~\cite{liu2023sctie}.}
	\label{fig:compare_d}
\end{figure}

\subsubsection{Scanning Performance}
\label{subsubsec:dualband_scan}

Fig.~\ref{fig:scan_cuts} shows the $\phi_0=0^\circ$ pattern cuts for $\theta_0=0^\circ$--$50^\circ$
in each band, obtained with a separate amplitude--phase distribution per band. In both bands the
main beam points at the commanded direction with no grating lobe.

% legend macros (move to the preamble)
\newcommand{\legitem}[3]{%
	\tikz[baseline=-0.6ex]{\draw[#1,line width=0.9pt] (0,0)--(1.0,0);
		\node[#1,fill=#1,#2,inner sep=0pt,minimum size=3pt] at (0.5,0) {};}%
	\,$\theta_0=#3^\circ$}

\newcommand{\cutlegend}{%
	\begin{tabular}{@{}l@{\hspace{1.2em}}l@{\hspace{1.2em}}l@{}}
		\legitem{c0}{circle}{0}  & \legitem{c1}{rectangle}{10} & \legitem{c2}{regular polygon,regular polygon sides=3}{20} \\[0.6ex]
		\legitem{c3}{diamond}{30} & \legitem{c4}{regular polygon,regular polygon sides=3,shape border rotate=180}{40} & \legitem{c5}{regular polygon,regular polygon sides=3,shape border rotate=270}{50}
\end{tabular}}

\begin{figure}[!t]
	\centering
	{\footnotesize\cutlegend}\\[1ex]
	\subfloat[]{\includegraphics[width=0.45\columnwidth]{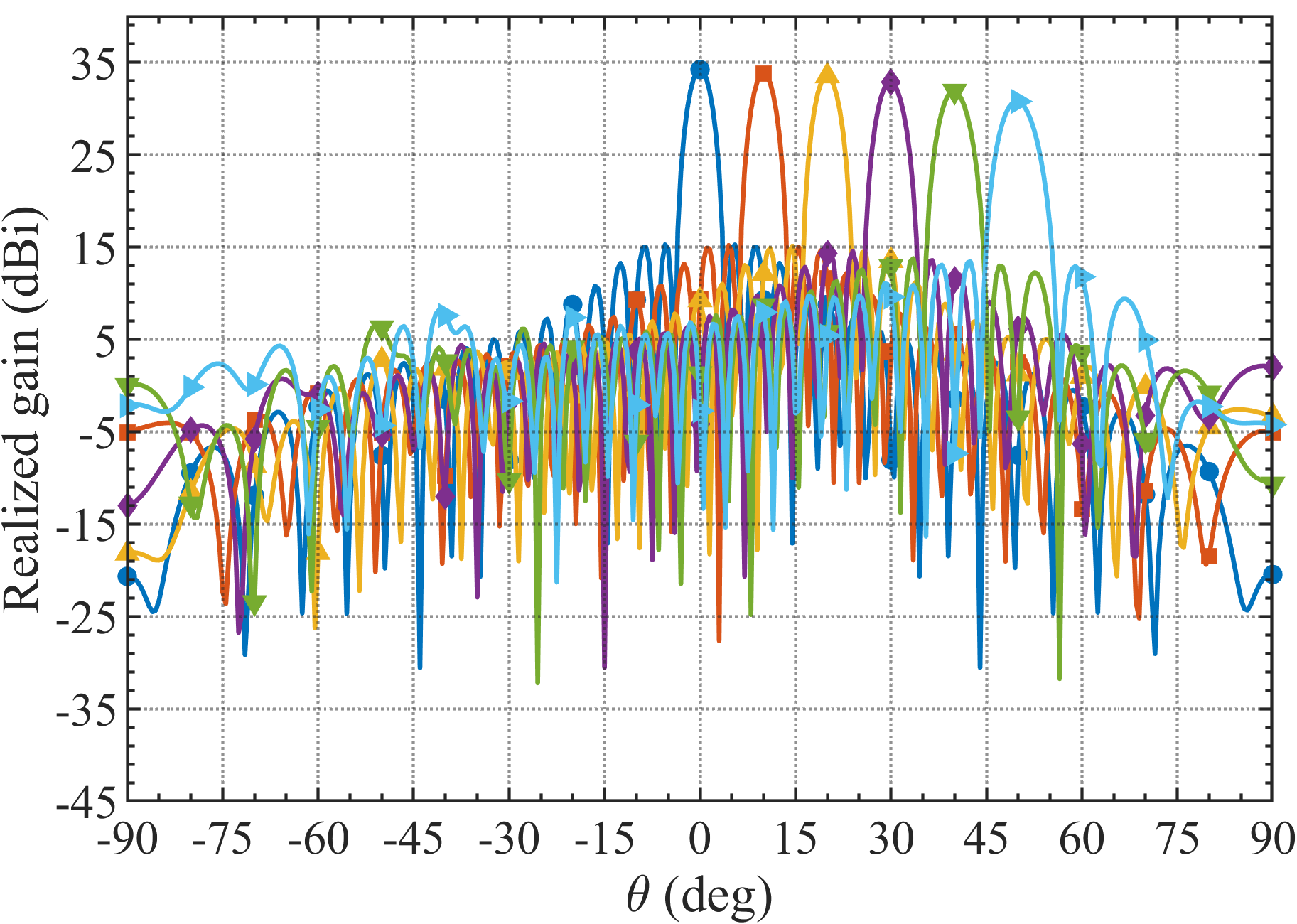}%
		\label{fig:cut_LL}}
	\hfil
	\subfloat[]{\includegraphics[width=0.45\columnwidth]{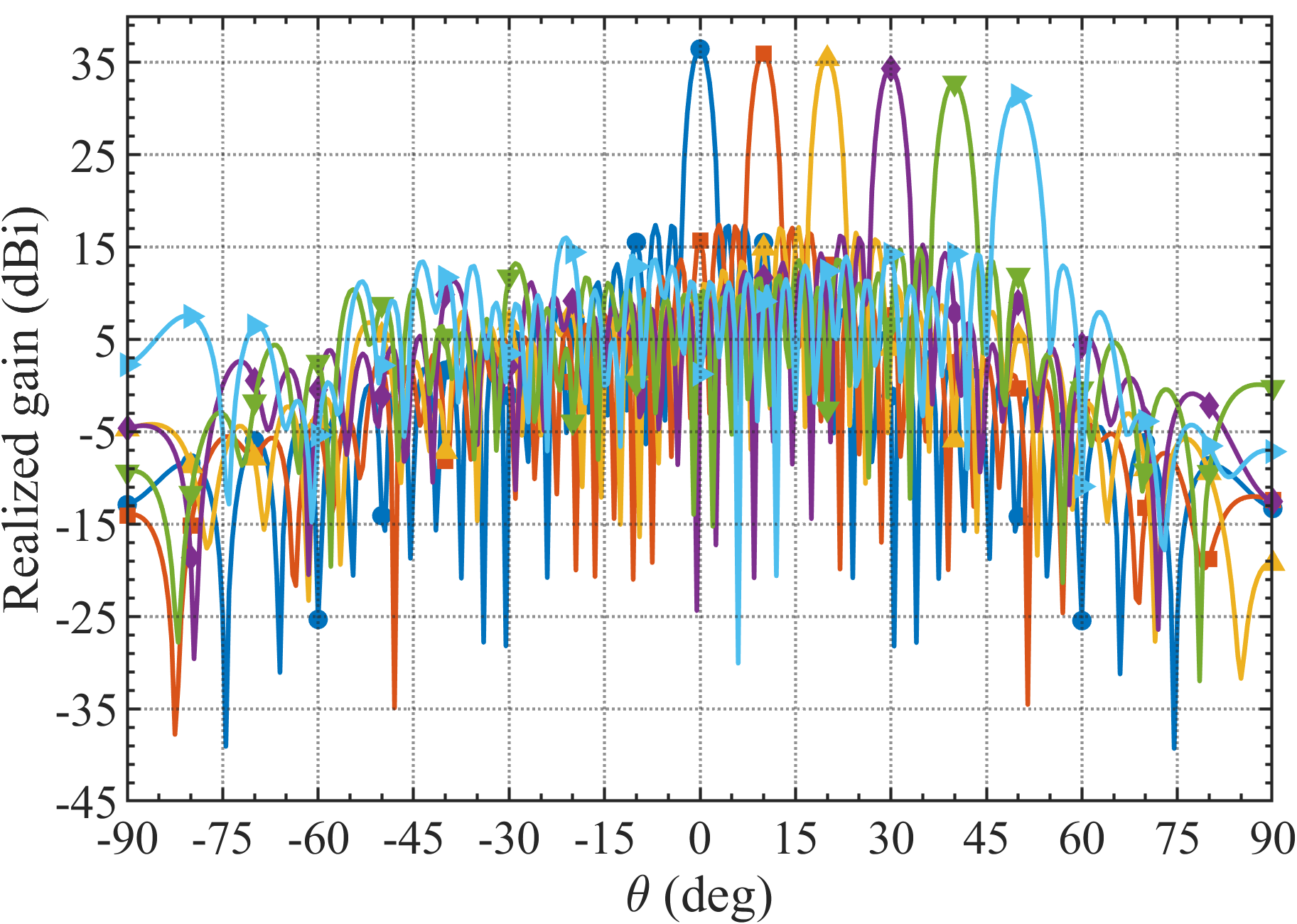}%
		\label{fig:cut_HH}}
	\caption{Realized gain pattern cuts for $\theta_0=0^\circ$--$50^\circ$ ($\phi_0=0^\circ$) with $d=11.71$~mm. (a)~11.2~GHz. (b)~14.0~GHz.}
	\label{fig:scan_cuts}
\end{figure}

Fig.~\ref{fig:gain_sll} shows the peak realized gain and peak SLL of the final design
($d=11.71$~mm) versus scan angle in the $\phi_0=0^\circ$ plane. At broadside, the peak realized
gain is 34.24~dBi at 11.2~GHz and 36.46~dBi at 14.0~GHz, decreasing monotonically to 30.79 and
31.40~dBi at $\theta_0=50^\circ$, i.e., scan losses of 3.45 and 5.06~dB. The upper band thus
degrades faster, and its realized-gain advantage over the lower band shrinks from 2.22~dB at
broadside to 0.61~dB at $50^\circ$. Part of the loss is the projected-aperture factor,
$-10\log_{10}\cos50^\circ=1.92$~dB. Since the LHCP gain at the beam peak is more than 20~dB below
the RHCP gain, the polarization loss is below 0.05~dB. The remaining 1.5~dB (lower band) and
3.1~dB (upper band) are most likely due to the roll-off of the element pattern (the half-power
beamwidths are $92^\circ$ and $76^\circ$), which is stronger at the higher frequency; because
mutual coupling is not included in the array model, any scan-dependent coupling contribution
remains to be verified.

The peak SLL of the lower band rises slowly from $-19.0$~dB at broadside to $-18.1$~dB at
$\theta_0=40^\circ$ and $-17.4$~dB at $50^\circ$. The upper band follows from $-19.1$~dB to about
$-18.0$~dB at $40^\circ$ but rises more sharply to $-15.2$~dB at $50^\circ$. The $-13.5$~dB target
is therefore met at every scan angle in both bands, with worst-case margins of 3.9~dB
(11.2~GHz) and 1.7~dB (14.0~GHz) at $50^\circ$. The optimization constrains the array factor,
whereas these SLL values are evaluated on the full patterns including the element pattern, so
the array-factor constraint carries over for these designs, although the margin at the largest
scan angle is limited. The sharper rise of the upper band at $50^\circ$ is consistent with its
stronger element-pattern roll-off, which lowers the main beam while the sidelobes are attenuated
less, and with its smaller grating-lobe margin.

\newcommand{\gslegitem}[3]{%
	\tikz[baseline=-0.6ex]{\draw[#1,line width=0.9pt] (0,0)--(1.0,0);
		\node[#1,fill=#1,#2,inner sep=0pt,minimum size=3pt] at (1.0,0) {};}%
	\,#3}

\newcommand{\gslegend}{%
	\begin{tabular}{@{}l@{\hspace{1.5em}}l@{}}
		\gslegitem{c0}{circle}{11.2~GHz} &
		\gslegitem{c1}{rectangle}{14.0~GHz}
\end{tabular}}

\begin{figure}[!t]
	\centering
	{\footnotesize\gslegend}\\[1ex]
	\subfloat[]{\includegraphics[width=0.45\columnwidth]{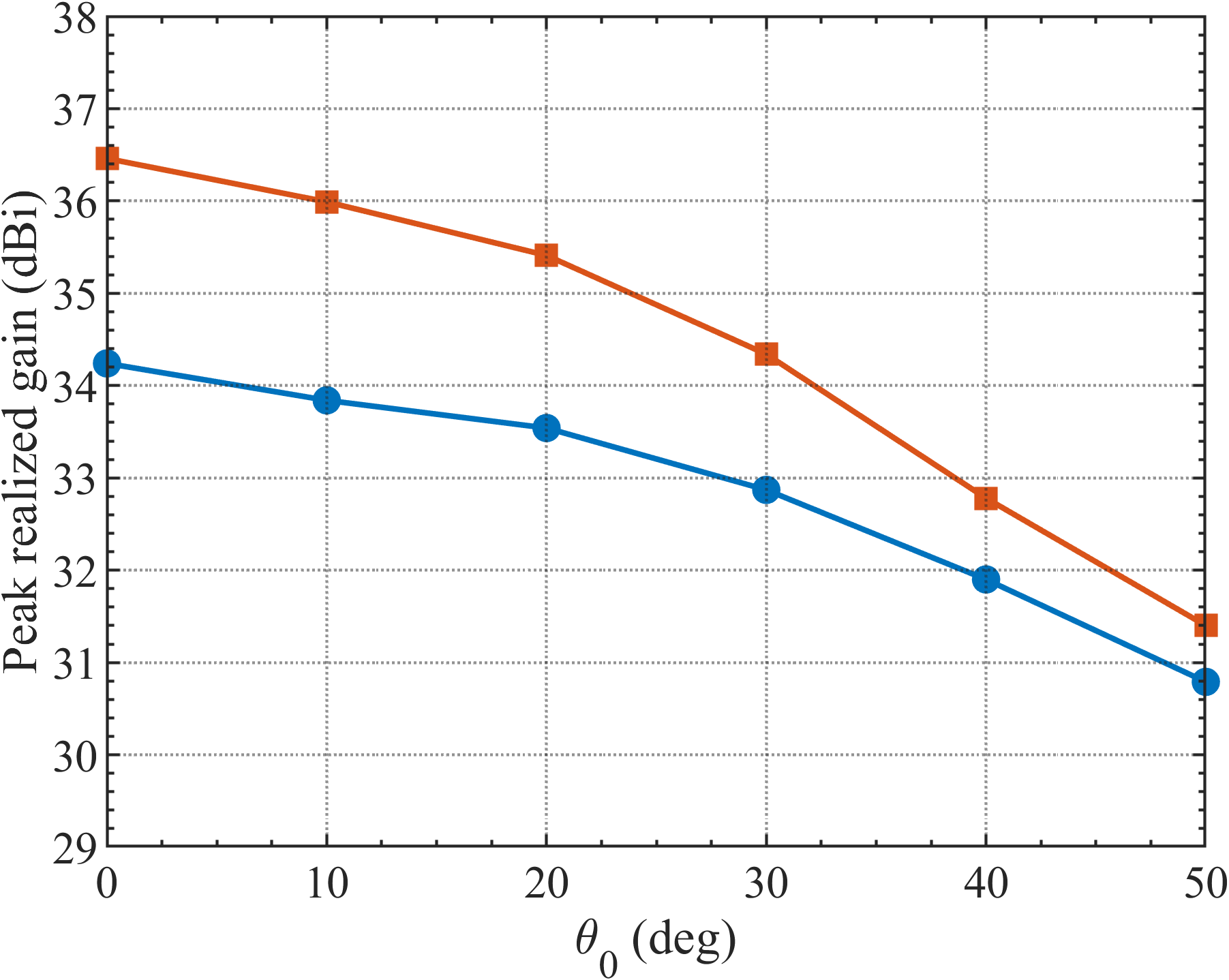}%
		\label{fig:gain_vs_theta}}
	\hfil
	\subfloat[]{\includegraphics[width=0.45\columnwidth]{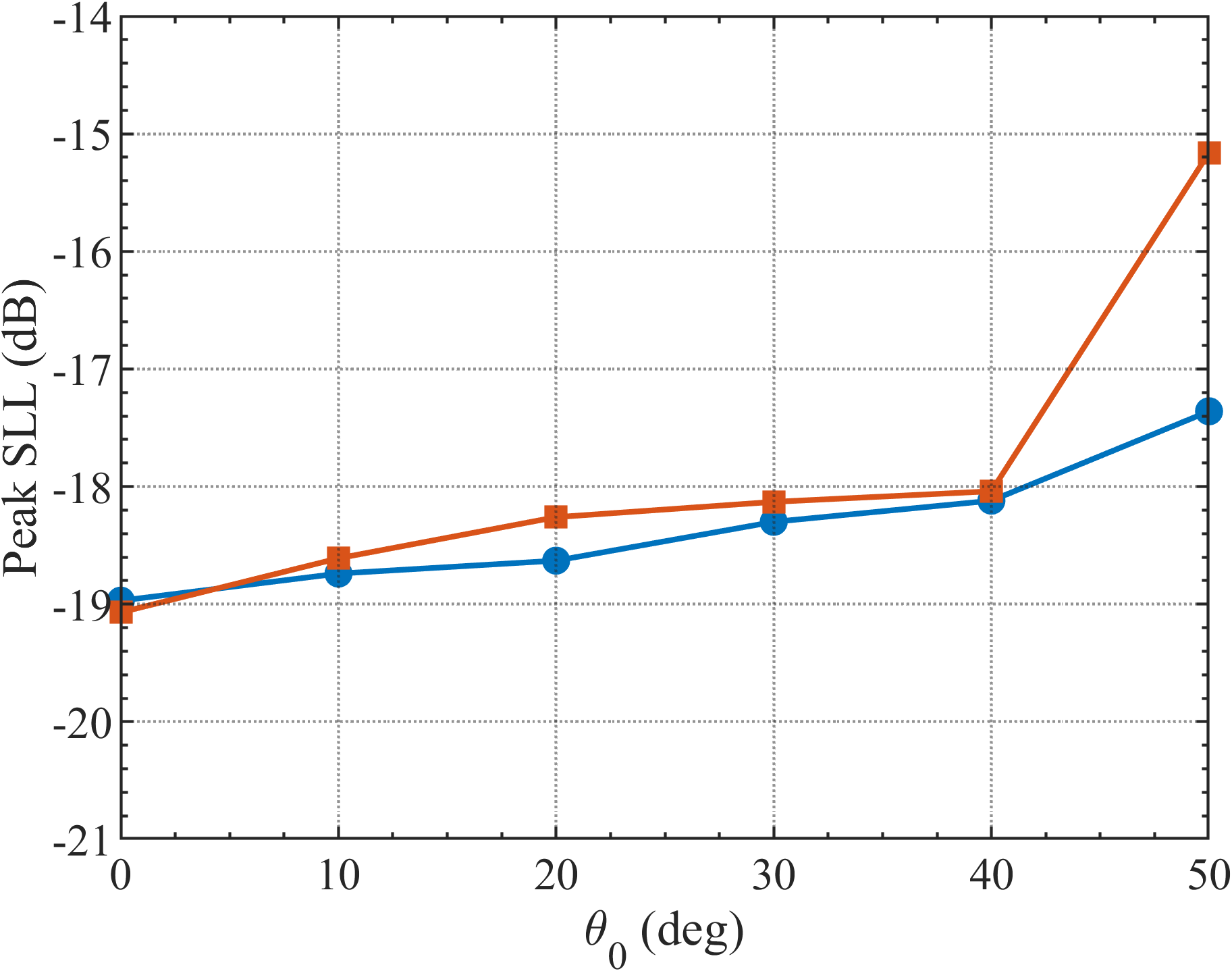}%
		\label{fig:sll_vs_theta}}
	\caption{(a)~Peak realized gain and (b)~peak SLL of the dual-band array versus scan angle $\theta_0$ ($\phi_0=0^\circ$) for $d=11.71$~mm.}
	\label{fig:gain_sll}
\end{figure}

\begin{figure}[!t]
	\centering
	\subfloat[]{\includegraphics[width=0.45\columnwidth]{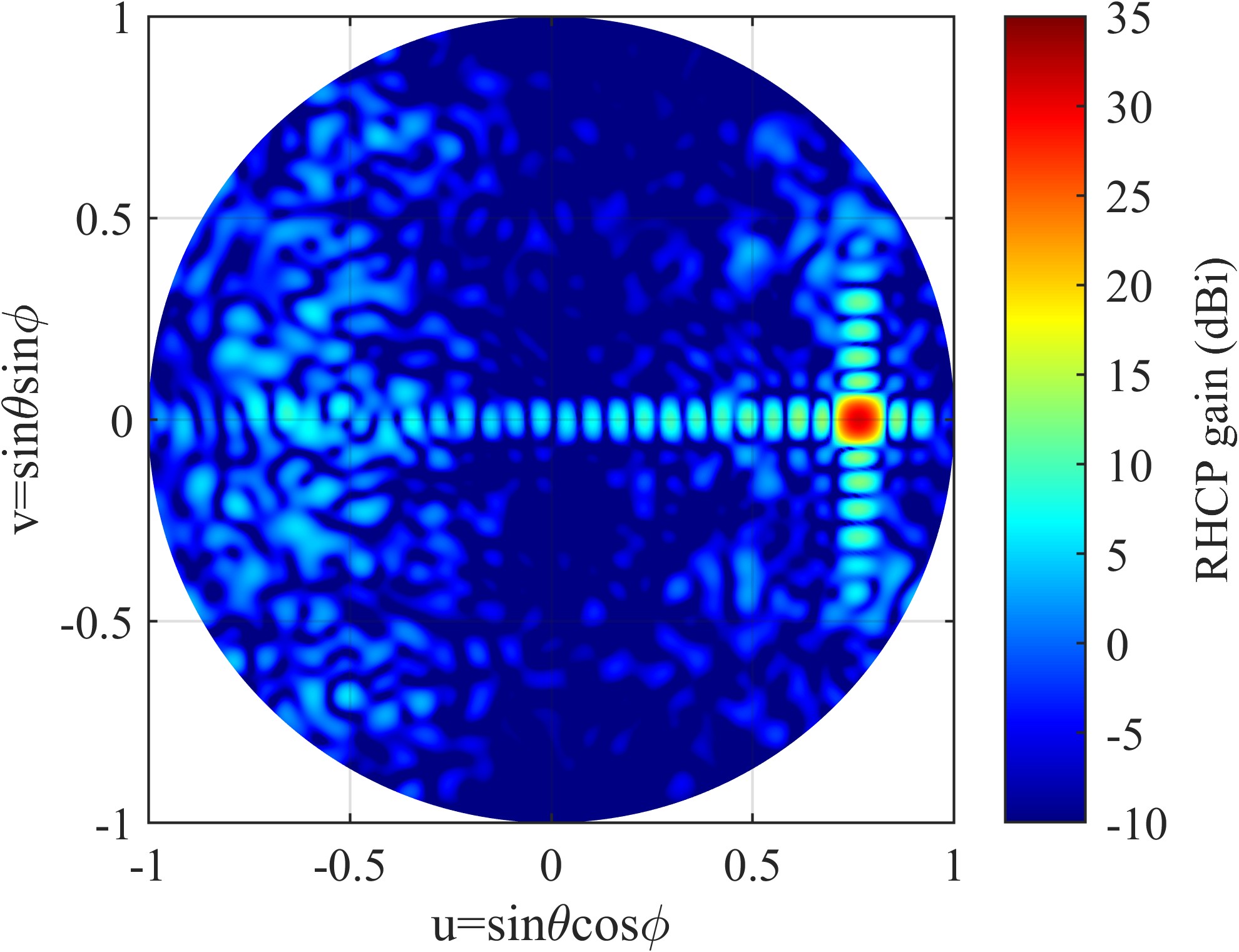}%
		\label{fig:uv_R_low}}
	\hfil
	\subfloat[]{\includegraphics[width=0.45\columnwidth]{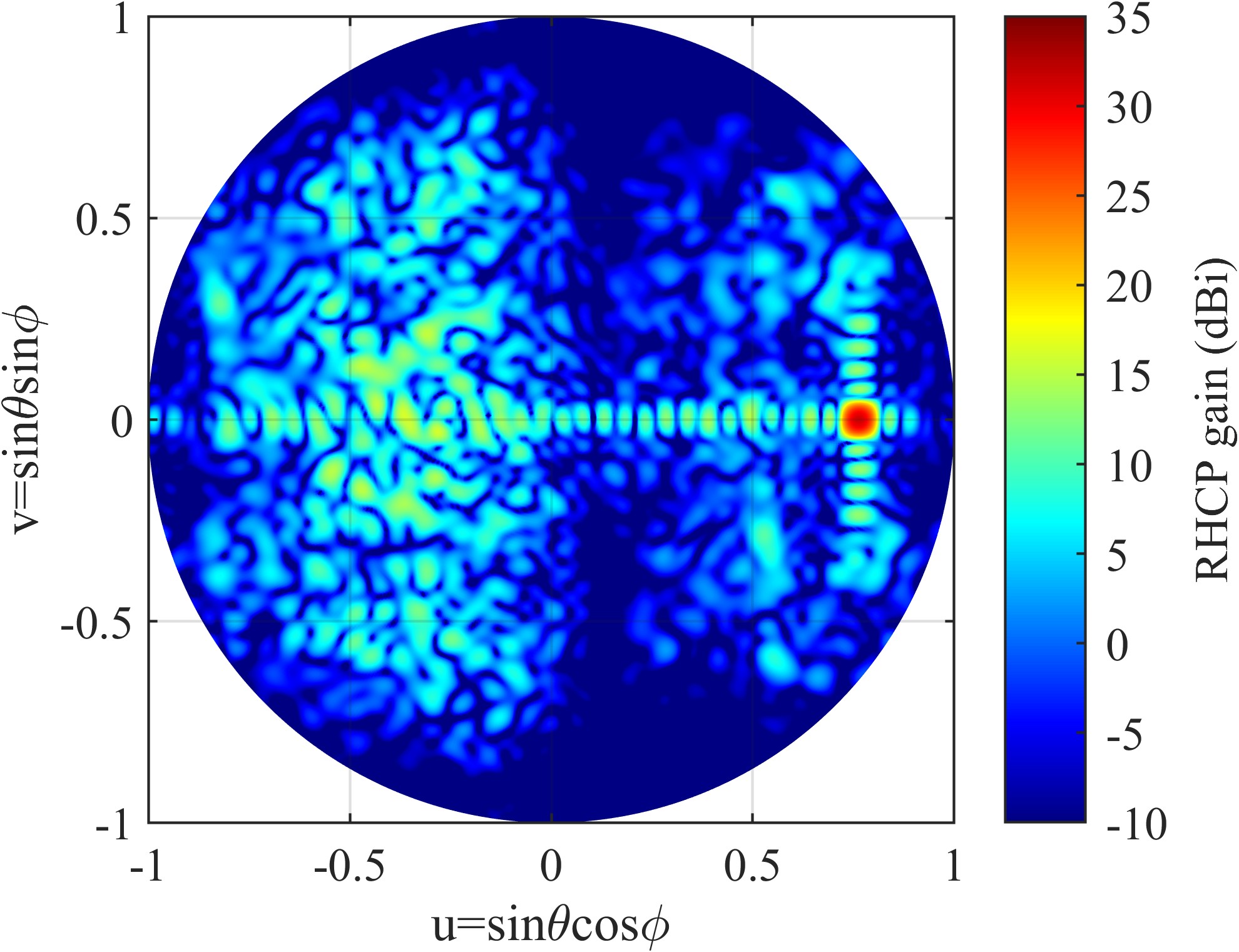}%
		\label{fig:uv_R_high}}
	\caption{RHCP radiation patterns in the $u$--$v$ plane of the optimized dual-band array scanned to $\theta_0=50^\circ$ ($\phi_0=0^\circ$) with $d=11.71$~mm. Each panel has its own color scale. (a)~11.2~GHz. (b)~14.0~GHz.}
	\label{fig:uv_patterns}
\end{figure}

Fig.~\ref{fig:uv_patterns} shows the RHCP $u$--$v$ patterns at $\theta_0=50^\circ$
($\phi_0=0^\circ$). In both bands the main beam appears at the intended position
$(u,v)=(\sin50^\circ,0)=(0.77,0)$, i.e., with no pointing error, and no grating lobe is observed
in the visible region, in agreement with the grating-lobe condition of the previous subsection
($d/\lambda$ equals 0.44 and 0.55 in the two bands). The margin is small at 14.0~GHz, where the
condition holds up to a scan angle of about $56^\circ$. The highest sidelobes lie along the two
principal cuts through the main beam. The LHCP component (not shown) is more than 27~dB below
the RHCP level at the beam peak and stays below $-10$~dBi elsewhere in the visible region, which
is consistent with the axial-ratio behavior of the element (Fig.~\ref{fig:AR_maps}).

In summary, the $36\times36$ domino-tiled array needs 648 control points per band and scans to
$50^\circ$ in both bands with no grating lobe, no pointing error, and a peak SLL of at most
$-17.4$~dB (11.2~GHz) and $-15.2$~dB (14.0~GHz), both below the $-13.5$~dB target. The upper band
at $50^\circ$ is the limiting case for both scan loss and SLL.

\section{Conclusion}
\label{sec:conclusion}

This paper presented a hybrid deep learning and optimization framework for the design of large irregular domino-tiled phased arrays. A local-flip Markov chain generator produces balanced domino tilings about $5.2\times$ faster than a randomized greedy baseline, and a Tabu Search engine achieves a lower SLL than a genetic algorithm, simulated annealing, and iterated local search at comparable directivity. These tools were used to build the dataset of a convolutional neural network (CNN) surrogate, which predicts from the tiling structure whether a candidate has a favorable multi-angle SLL. For apertures from $20\times20$ to $100\times100$ elements, using it as a gate in the Tabu Search reduced the full evaluations by $60.6$--$70.6\%$ and the search time by a factor of $2.56$--$3.80$ with equal or slightly higher fitness, and a single run including the one-time costs was faster for apertures of $50\times50$ and larger. A convex amplitude--phase optimization of the fixed tilings then gave equal or higher directivity than two state-of-the-art methods at a comparable sidelobe level, with more than an order of magnitude shorter computation time, and it compensated for a tiling not matched to the scan angle.

The framework was also applied to a $36\times36$ dual-band circularly polarized array for LEO terminals at $11.2$ and $14.0$~GHz, with $648$ dominoes and a common element spacing selected from a sweep. The array scans up to $50^\circ$ with no grating lobe and, compared with a reference irregular array, gives higher directivity over most of the scan range, a lower SLL at large scan angles, and a more practical spacing for the lower band. Each band needs its own amplitude--phase distribution.

Future work will extend the convex optimization to low-resolution phase shifters and dual-polarized elements, optimize the tiling jointly for both bands, validate the dual-band array with full-wave simulation including mutual coupling and a measured prototype, and apply the framework to large arrays for 5G/6G and sub-terahertz systems.

\IEEEtriggeratref{80}
\IEEEtriggercmd{\newpage}

%\vspace*{-5\baselineskip}
\vspace{-1.5cm}   
\begin{IEEEbiography}[{\includegraphics[width=1in,height=1.25in,clip,keepaspectratio]{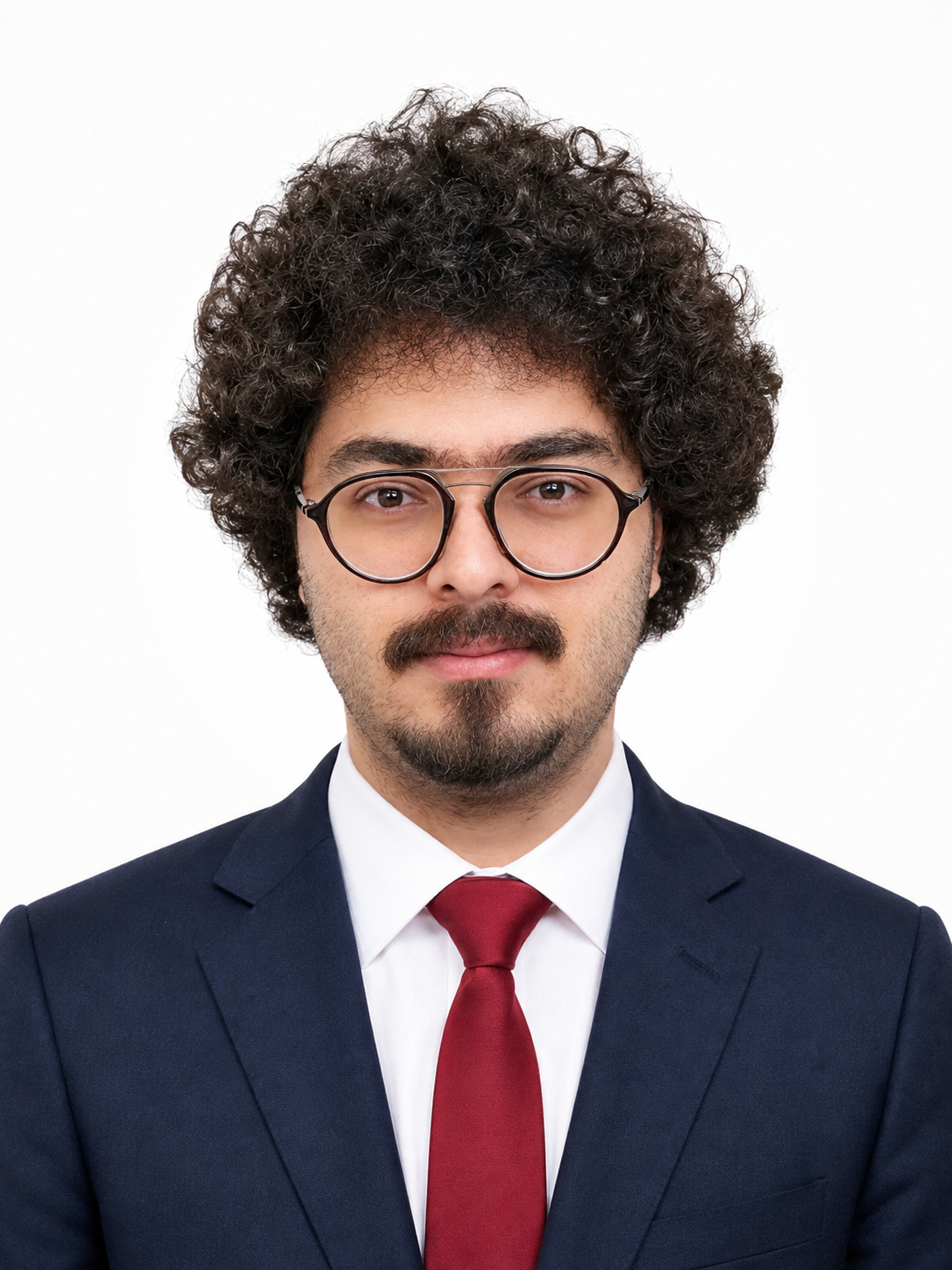}}]
{Mohammad Mashayekhi} received the B.Sc. degree in electrical engineering from the University of Kashan, Kashan, Iran, in 2019, and the M.Sc. degree in electrical engineering from the Iran University of Science and Technology (IUST), Tehran, Iran, in 2023. Since 2023, he has been a Researcher with the Information and Communication Technology Institute, Isfahan University of Technology (IUT), Isfahan, Iran. His research interests include antenna engineering, metasurfaces, antenna arrays, inverse electromagnetic design, computational electromagnetics, and the application of deep learning and machine learning in electromagnetics.
\end{IEEEbiography}

%\vspace*{-2\baselineskip}
\vspace{-1.5cm}   
\begin{IEEEbiography}[{\includegraphics[width=1in,height=1.25in,clip,keepaspectratio]{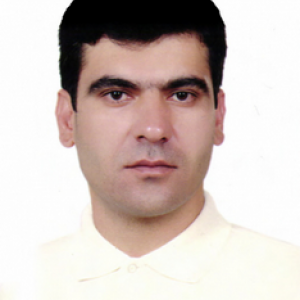}}]{Gholamreza Askari}
	received the B.Sc. and M.Sc. degrees in electrical engineering from Isfahan University of Technology, Isfahan, Iran, in 1992 and 1994, respectively, and the Ph.D. degree in electrical engineering (fields and waves) from the University of Tehran, Tehran, Iran, in 2017. He is currently an Assistant Professor at the Information and Communication Technology Institute (ICTI). His research interests include antenna arrays, electromagnetic theory, microwave engineering, and computational electromagnetics.
\end{IEEEbiography}

\end{document}